%% file: variations.tex
\newcommand\relativepath{}

\documentclass[a4paper,11pt]{article}
\usepackage{fullpage}

\newcommand\bigone[1]{}
\newcommand\smallone[1]{#1}
\input{\relativepath _paper}

\usepackage{algorithm}

\newcommand\draft[1]{}
\newcommand\release[1]{#1}
\release{\usepackage[breaklinks]{hyperref}}

\usepackage{enumerate}

\usepackage{euscript}
\newcommand{\circuit}[1]{\EuScript{#1}}

\def \ket#1|#2>%
{\ifx&#1& 
|#2\rangle
\else
|#2\rangle_{\mathcal #1}
\fi}

\newcommand{\Adv}{\mathop{\mathrm{ADV}^{\pm}}}

\begin{document}

\title{Variations on Quantum Adversary}
\author{Aleksandrs Belovs\thanks{
Faculty of Computing, University of Latvia.}}
\date{}
\maketitle

\begin{abstract}
The (negative-weighted) quantum adversary bound is a tight characterisation of the quantum query complexity for any partial function~\cite{hoyer:advNegative, reichardt:advTight}.  
We analyse the extent to which this bound can be generalised.
Ambainis \etal~\cite{ambainis:symmetryAssisted} and Lee \etal~\cite{lee:stateConversion} generalised this bound to the state generation and state conversion problems, respectively.
Using the ideas by Lee \etal, we get even further generalisations of the bound.

We obtain a version of the bound for general input oracles, which are just arbitrary unitaries.  We also generalise the bound to the problem of implementing arbitrary unitary transformations.  Similarly to the bound by Lee~\etal, our bound is a lower bound for exact transformation and an upper bound for approximate transformation.
This version of the bound possesses the tight composition property.

Using this construction, we also obtain lower bounds on the quantum query complexity of functions and relations with general input oracles.
\end{abstract}

\section{Introduction}

General (negative-weighted) adversary method is one of the most interesting developments in quantum query algorithms.  This is a concise semidefinite optimisation problem that tightly (up to a constant factor) characterises the bounded-error quantum query complexity of evaluating any function $f\colon \cD\to [m]$ with $\cD\subseteq[q]^n$.

The adversary bound originates from the hybrid method, which was used by Bennett \etal~\cite{bennett:strengths} to show a tight $\Omega(\sqrt{n})$ lower bound on quantum search.
The hybrid method proceeds by analysing the weights with which the quantum algorithm queries different variables for various inputs, and finds a pair of inputs that the algorithm fails to distinguish.

This idea was further refined by Ambainis in the first (\emph{unweighted}) version of the adversary bound~\cite{ambainis:adv}.  Instead of searching for one pair of inputs, the adversary method studies many pairs and shows that any algorithm has to make many queries in order to distinguish all of them simultaneously.
The unweighted adversary bound is more powerful than the hybrid method, and it is easy to use due to its attractive combinatorial formulation.  This resulted in a large number of applications: \cite{Durr:quantumGraph, berzina:graphProblems, buhrman:productVerification, dorn:algebraicProperties} to name a few.
Soon afterwards came the \emph{positive-weighted} version of the bound~\cite{ambainis:polVsQCC, zhang:advPower}.  In 2004 \v Spalek and Szegedy~\cite{spalek:advEquivalent} mentioned 7 different versions of the bound and proved they all are equivalent.

The next important step was done by H\o yer \etal~\cite{hoyer:advNegative}.
Starting with the semidefinite formulation of the positive-weighted adversary bound by Barnum \etal~\cite{barnum:advSpectral}, they showed that the same expression still yields a lower bound if one replaces non-negative entries by arbitrary real numbers.  This \emph{negative-weighted} version of the bound is strictly more powerful than the positive-weighted one, but it is also harder to apply.

In a series of papers~\cite{reichardt:formulae, reichardt:spanPrograms, reichardt:advTight}, Reichardt \etal surprisingly proved that the negative-weighted version of the bound is tight:  The dual formulation of the bound (which is equal to the primal formulation due to strong duality) can be transformed into a quantum query algorithm with the same complexity!
Thus, instead of devising a quantum query algorithm, one may come up with a feasible solution to the dual adversary bound.

The negative-weighted adversary bound has been used to prove lower bounds~\cite{spalek:kSumLower, belovs:onThePower, belovs:setEquality}, but more frequently to prove upper bounds, in particular using the learning graph approach~\cite{belovs:learning}.  For instance, the adversary bound was used to construct quantum algorithms for formula evaluation~\cite{reichardt:formulae,  reichardt:unbalancedFormulas, zhan:treesWithHiddenStructure}, finding subgraphs~\cite{lee:learningTriangle, belovs:learningClaws, legall:constSizedHypergraphs}, $k$-distinctness problem~\cite{belovs:learningKDist}, and in learning and property testing~\cite{belovs:learningSymmetricJuntas, belovs:monotonicityQuantum}.

One of the main features of the dual adversary bound is that it provides an ``idealised'' analogue of a quantum query algorithm.  Its complexity equals the complexity of a bounded-error algorithm, but it behaves as if it were error-free:
Feasible solutions for a function $f$ and a function $g$ can be combined into a feasible solution for the composed function $f\circ g$, whose complexity is the product of the two.  This is known as the \emph{tight composition property}, and it is at the heart of the formula evaluation algorithm by Reichardt and \v Spalek~\cite{reichardt:formulae}.  Thus, for a fixed Boolean function $f$, evaluation of the quantum query complexity of the composed function $f^{d}$ is a routine computation now.
Let us note that no classical counterpart to this construction is known.  For instance, the  randomised query complexity of $f^{d}$ is still unknown even when $f$ is the majority function on 3 bits~\cite{magniez:maj3, leonardos:maj3}.

The adversary bound was later generalised to the \emph{state generation} problem by Ambainis \etal~\cite{ambainis:symmetryAssisted}, and to the \emph{state conversion} problem by Lee \etal~\cite{lee:stateConversion}.  In the latter problem, one is given oracle access to a string $x\in[q]^n$, and the task is to transform the \emph{initial state} $\rho_x$ into the \emph{target state} $\sigma_x$ for all $x$.
This generalisation comes with additional subtleties.
The complexity of evaluating a Boolean function $f$ has a jump at error $\eps = 1/2$: It is trivial for $\eps = 1/2$, but becomes complicated for any smaller error parameter.  State generation problem can have various jumps even at small values of $\eps$ (for example, consider the state generation problem with the target states $\sigma_x = \sqrt{1-\eps}\ket A|0> + \sqrt{\eps} \ket B|f(x)>$ in the space $\cA \oplus \cB$).
The adversary bound, whose value for each problem is just one real number, cannot account for all these jumps, and, consequently, cannot be tight for all values of $\eps$.
Lee \etal bypassed this complication by constructing a bound that is \emph{semi-tight}: 
It is still an upper bound for \emph{approximate} transformation, but it is a lower bound only for \emph{exact} transformation.
A lower bound for approximate transformation can be obtained by minimising the semi-tight bound over all target states that are sufficiently close to the exact target state $\sigma_x$.
The semi-tight bound for state conversion is stated in terms of the \emph{filtered $\gamma_2$-norm}, which is a generalisation of the usual $\gamma_2$-norm.

Besides this, the paper by Lee \etal~\cite{lee:stateConversion} has a number of additional important ideas.  
The proof of the upper bound was significantly simplified by introducing easy and powerful Effective Spectral Gap Lemma, which can be also used independently~\cite{belovs:electicityQuantumWalks, belovs:mergedWalk3Dist}.
Next, unlike~\cite{hoyer:advNegative}, the lower bound is proven directly in the dual form, essentially showing that the adversary bound is a semidefinite relaxation of a quantum query algorithm.
This proof is not only shorter, but also gives an algorithmic interpretation of the coefficients of the dual adversary bound.

\paragraph{Main Results}
The aim of this paper is to give a more general and refined treatment of the ideas introduced by Lee \etal in~\cite{lee:stateConversion}, and to analyse some of their consequences.

First, all of the aforementioned variants of the adversary bound only work for a very restricted variant of the input oracle, which we call the \emph{standard} input oracle.  It is of the form $\ket |j>\ket |0>\mapsto \ket |j>\ket |x_j>$, where $x = (x_j)\in [q]^n$ is the input string.
We show that the same results hold if we replace this restricted oracle by a \emph{general} input oracle, which is just an arbitrary unitary transformation.
This highlights the essence of the construction by removing the details arising from the special case of the input oracle.
General oracles naturally appear in applications.
For instance, the well-known phase detection procedure (see \rf(thm:phaseDetection) below) is clearly of this form.

Second, we generalise the adversary bound from the state conversion problem to the problem of implementing arbitrary unitary transformations.
This problem has also appeared in applications, as an example, we can mention simulation of sparse Hamiltonians, see~\cite{berry:exponentialSimulatingHamiltonians} and references therein.

Our version of the adversary bound also has the tight composition property.  We prove it for composition of procedures, as well as for direct sums and tensor products.
This variant of the tight composition property can be helpful in applications, for instance, in nested quantum walks~\cite{jeffery:nestedWalks}.
This property also facilitates modular approach, in which parts of the construction, like the input and the output of the algorithm, can be studied independently and combined in various ways.

Finally, we use these general tools to give lower bounds on the quantum query complexity of functions and relations (problems with many possible outputs for each input).  Up to our knowledge, the case of relations has not been studied previously even for the standard input oracle.
We prove that the adversary bound still gives a tight lower bound for bounded-error function evaluation with general input oracles.
We also show a similar result for evaluation of relations, provided that the output of the algorithm can be efficiently tested.

\paragraph{Outline and Other Results}
Our version of the adversary bound is based on another generalisation of the $\gamma_2$-norm, which we call \emph{relative $\gamma_2$-norm}.
We define the relative $\gamma_2$-norm and prove a number of its properties in \rf(sec:gamma2).  
With this notion, our adversary bounds for state conversion and unitary implementation attain a very concise form.

The proof of the semi-tightness of our adversary bound for state conversion in \rf(sec:conversion) is essentially a refined version of the corresponding proofs by Lee \etal~\cite{lee:stateConversion}.
The bound for the unitary implementation problem in \rf(sec:unitary) is genuinely new, and proceeds by a simple reduction to state conversion.  The tight composition results easily follow from the corresponding properties of the relative $\gamma_2$-norm.

In \rf(sec:fractional), we formulate a version of our adversary bound for the \emph{fractional} query complexity model.  The latter is equivalent~\cite{cleve:discreteTimeSimulation} to the continuous-time query model\footnote{Strictly speaking, this was only shown for the standard input oracle, but the corresponding proof easily carries to the general case.}, which is a well-known competitor of the discrete-time query model~\cite{farhi:analog}.
For standard input oracles, the continuous- and discrete-time query models are known to be equivalent~\cite{yonge-mallo:advForContinuous, lee:stateConversion}.  Curiously, this result was shown using the adversary bound.  For general oracles, however, the corresponding query complexities can differ substantially.  

In \rf(sec:preparation), we study input oracles that generate quantum states.
This is an intermediate step between general and standard input oracles.
It is reasonable to assume that subroutines of a quantum algorithm can communicate in quantum states, not only classical messages, so this model is worth studying.
For example, Eisentr\"ager \etal~\cite{eisentrager:unitGroup} gave a very nice treatment of the continuous HSP with input oracles that generate quantum states.
We give two versions of the bound, one easier to use, but not tight, and another one semi-tight, but harder to use.  In \rf(sec:counterexample), we compare the performance of both bounds on a simple problem: exact amplitude amplification.  
In addition, we show that the adversary bound for unitary implementation is more precise than the adversary bound for state conversion.

Lower bounds on quantum query complexity of relations in \rf(sec:relationBasic) are obtained by optimising the adversary bound over all possible target states.
In \rf(sec:purifiers), we show that the adversary bound tightly characterises bounded-error quantum query complexity of functions.  
In the spirit of modular approach, we prove this by constructing an explicit solution to the adversary bound for the ``purification'' problem:  Given an oracle that generates a noisy state, generate the corresponding exact state.

We end the paper with a number of open problems.


\section{Preliminaries}



If not said otherwise, a \emph{vector space} is a finite-dimensional complex inner product space.  They are denoted by calligraphic letters.  We assume that each vector space has a fixed orthonormal basis, and we often identify an operator with the corresponding matrix.
The inner product is denoted by $\ip<\cdot,\cdot>$.
$A^*$ stands for the adjoint linear operator, and $A\elem[i,j]$ for the $(i,j)$th entry of the matrix $A$.
$A\circ B$ stands for the Hadamard (entry-wise) product of matrices.
$I_{\cX}$ stands for the identity operator in $\cX$.
All projectors are orthogonal projectors.
We often treat scalars as $1\times1$ matrices.

For a linear operator $A\colon \cX_2\to\cX_1$, we have the \emph{singular value decomposition} $A = \sum_{i=1}^k \sigma_i u_i v_i^*$, where $k$ is the rank of $A$, $\sigma_i$ are positive reals, $\{u_i\}$ is an orthonormal family (not necessarily complete) of vectors  in $\cX_1$, and $\{v_i\}$ is an orthonormal family in $\cX_2$.  $\sigma_i$, $u_i$ and $v_i$ are called the \emph{singular values, left singular vectors} and \emph{right singular vectors} of $A$, respectively.  The \emph{spectral norm} of $A$ is $\|A\| = \max_i \sigma_i$, and the \emph{trace norm} is $\normtr|A| = \sum_i \sigma_i$.  

A \emph{quantum register} is identified with the corresponding vector space.  
A \emph{quantum state} is a unit vector in this space (we only consider pure states).
If two vector spaces $\cX$ and $\cY$ are isomorphic, we assume that there is an implicit isomorphism between the two.  If $\cY$ is additionally a register and $\psi\in\cX$, we use notation $\ket Y|\psi>$ to denote the image of $\psi$ under this isomorphism.  This is to help to keep track to which of the isomorphic registers the vector belongs.
If $\{e_j\}$ is the standard orthonormal basis of $\cX$, we use a shorthand $\ket X|j>$ instead of $\ket X|e_j>$.

We use $1_P$ to denote the indicator variable, that equals 1 if $P$ is true, and equals 0 otherwise.


\paragraph{Quantum Query Algorithms}
Let $\cX$ be a vector space.  A \emph{quantum algorithm with oracle} $O$ acting on $\cX$ has the state space of the form $\cS\oplus (\cX\otimes\cW)$.  The algorithm is a unitary transformation of the form
\begin{equation}
\label{eqn:algorithm}
U_0 \to \tO^{\pm1} \to U_1\to \tO^{\pm1}\to \cdots \to U_{T-1} \to \tO^{\pm1}\to U_T,
\end{equation}
where $U_i$ are arbitrary unitaries on $\cS\oplus (\cX\otimes\cW)$ independent of $O$, and each $\tO^{\pm1}$ is a query to the input oracle.
It can be either a direct query, $\tO = I_\cS + O\otimes I_\cW$, or a reverse query, $\tO^{-1} = I_\cS + O^*\otimes I_\cW$.
Using standard techniques, any quantum query algorithm can be converted into this form. 
The oracle $O$ should be interpreted as the input of the algorithm, and we often call it the \emph{input oracle}.  
It can be replaced by an arbitrary unitary on $\cX$, and, for different choices, the algorithm implements a different unitary.
The number $T$ is the \emph{query complexity} of the algorithm, and our objective is to minimise it.

\paragraph{Technical Results}
The following two lemmata are used in the proof \rf(thm:conversionUpper).

\begin{lem}[Effective Spectral Gap Lemma~\cite{lee:stateConversion}] 
\label{lem:effective}
Let $\Pi_1$ and $\Pi_2$ be two projectors in the same vector space, and $R_1 = 2\Pi_1-I$ and $R_2 = 2\Pi_2-I$ be the reflections about their images.
For $\delta \ge 0$, let $P_\delta$ be the projector onto the span of all eigenvectors of $R_2R_1$ that have eigenvalues $\ee^{\ii\theta}$ with $|\theta|\le \delta$.  Then, for any vector $w$ in the kernel of $\Pi_1$, we have
\[ \|P_\delta \Pi_2 w \|\le \frac{\delta}{2}\|w\|. \]
\end{lem}

The following algorithm approximately reflects about the eigenvalue 1 eigenspace of the input oracle.
It is an easy consequence of the phase estimation algorithm.

\begin{thm}[Phase Detection~\cite{kitaev:phaseEstimation, cleve:phaseEstimation}]
\label{thm:phaseDetection}
For any vector space $\cX$, precision $\delta>0$, and error $\eps>0$, there exists a quantum algorithm with an oracle $O$ on $\cX$ satisfying the following properties:
\negmedskip
\itemstart
\item For each unit $\psi\in\cX$ satisfying $O\psi = \psi$, the algorithm leaves $\psi$ intact.
\item For each unit $\psi\in\cX$ satisfying $P_\delta\psi = 0$, the algorithm maps $\psi$ into a state $\phi$ satisfying $\|\psi+\phi\|\le\eps$.  
Here $P_\delta$ denotes the projector onto the space spanned by the eigenvectors of $O$ having eigenvalues $\ee^{\ii\theta}$ with $|\theta|\le \delta$.
\item The algorithm uses $O(\frac1\delta\log\frac1\eps)$ queries.
\itemend
\end{thm}

\section{Relative \texorpdfstring{$\gamma_2$}{gamma2}-norm}
\label{sec:gamma2}
In this section, we define the relative $\gamma_2$-norm, and prove some of its properties.
This is a generalisation of the usual and filtered $\gamma_2$-norms.
Most of our results in the paper are stated in terms of this bound.

\begin{defn}[Relative $\gamma_2$-norm]
\label{defn:matrixGamma2}
Let $\cX_1$, $\cX_2$, $\cZ_1$ and $\cZ_2$ be vector spaces, and $\cD_1$ and $\cD_2$ be sets of labels.
Let $A = \{A_{xy}\}$ and $\Delta = \{\Delta_{xy}\}$, where $x\in\cD_1$ and $y\in\cD_2$, be two families of linear operators: $A_{xy}\colon \cZ_2\to\cZ_1$ and $\Delta_{xy}\colon \cX_2\to\cX_1$.
The \emph{relative $\gamma_2$-norm}, 
\[
\gamma_2(A | \Delta) = \gamma_2(A_{xy} \mid \Delta_{xy})_{x\in\cD_1,\; y\in\cD_2},
\]
is defined as the optimal value of the following optimisation problem, where $\Upsilon_x$ and $\Phi_y$ are linear operators,
\begin{subequations}
\label{eqn:matrixGamma2}
\begin{alignat}{2}
&\mbox{\rm minimise} &\quad& \max \sfigB{ \max\nolimits_{x\in \cD_1} \norm|\Upsilon_x|^2, \max\nolimits_{y\in\cD_2} \norm|\Phi_y|^2 } \\
& \mbox{\rm subject to}&&  
A_{xy} = \Upsilon_x^* (\Delta_{xy}\otimes I_{\cW}) \Phi_y \qquad\qquad \text{\rm for all $x\in\cD_1$ and $y\in\cD_2$;}  \label{eqn:matrixGamma2Condition}\\
&&& \text{$\cW$ is a vector space,\quad 
$\Upsilon_x\colon \cZ_1\to \cX_1\otimes\cW$,\quad $\Phi_y\colon \cZ_2\to \cX_2\otimes\cW$.}
\end{alignat}
\end{subequations}
\end{defn}

This is a generalisation of the usual $\gamma_2$-norm, also known as Schur (Hadamard) product operator norm~\cite{bhatia:positive}.
Namely, for a $\cD_1\times \cD_2$ matrix $A$, the $\gamma_2$-norm of $A$ is defined by
\[
\gamma_2(A) = \gamma_2\sA[A{\elem[x,y]}\mid 1]_{x\in\cD_1,y\in\cD_2},
\]
with all $A_{xy}$ and $\Delta_{xy}$ being one-dimensional.
Moreover, if all $A_{xy} = a_{xy}$ and $\Delta_{xy}=\delta_{xy}$ are one-dimensional, and all $\delta_{xy}$ are non-zero, then the relative $\gamma_2$-norm can be reduced to the usual one:
\begin{equation}
\label{eqn:1dgamma2}
\gamma_2(a_{xy} \mid \delta_{xy})_{x\in\cD_1,y\in\cD_2} = \gamma_2\s[ \frac{a_{xy}}{\delta_{xy}}]_{x\in\cD_1,y\in\cD_2}.
\end{equation}

The filtered $\gamma_2$-norm of~\cite{lee:stateConversion} can be also expressed in this form.
Let $Z = \{Z_1,\dots,Z_n\}$ be a family of $\cD_1\times \cD_2$ matrices.  Then, the filtered $\gamma_2$-norm of $A$ with respect to $Z$ can be defined as $\gamma_2(A\elem[x,y]\mid\Delta_{xy})_{x\in\cD_1,y\in\cD_2}$, where all $\Delta_{xy}$ are diagonal $n\times n$ matrices given by $\Delta_{xy}\elem[i,i] = Z_i\elem[x,y]$.
We will not explicitly use this definition in our paper.


\begin{prp}
\label{prp:gamma2norm}
For a fixed family $\Delta$, $\gamma_2(A|\Delta)$ is finite on the linear space of families of matrices $A=\{A_{xy}\}$ satisfying $A_{xy} = 0$ whenever $\Delta_{xy}=0$.
Moreover, it is a vector norm on this space, that is, for all families $A$ and $B$ satisfying $\gamma_2(A |\Delta),\gamma_2(B|\Delta)<+\infty$, and any complex $c$, we have
\begin{enumerate}[(a)]
\itemsep0pt
\item \label{positive@prp:gamma2norm} positivity: 
$\gamma_2(A|\Delta)\ge 0$ with $\gamma_2(A|\Delta)=0$ iff $A=0$;

\item \label{homogen@prp:gamma2norm} homogeneity: 
$\gamma_2(c A_{xy}\mid \Delta_{xy})_{x\in\cD_1,y\in\cD_2} = |c|\, \gamma_2(A|\Delta)$; and

\item \label{triangle@prp:gamma2norm}triangle inequality: 
$\gamma_2(A_{xy}+B_{xy}\mid \Delta_{xy})_{x\in\cD_1,y\in\cD_2} \le \gamma_2(A|\Delta) + \gamma_2(B|\Delta)$.
\end{enumerate}
In particular, this norm is continuous when restricted to this space.
The triangle inequality is also valid for infinite sums and integrals.
\end{prp}

\pfstart
Property (a) trivial.
Next, let $\{\Upsilon_x\}, \{\Phi_y\}$ and $\{\Upsilon'_x\}, \{\Phi'_y\}$ be optimal solutions to~\rf(eqn:matrixGamma2) for the matrices $A$ and $B$ with objective values $W$ and $W'$, respectively.  Then, $\{\sqrt{|c|}\, \Upsilon_x\}$ and $\sfigA{\frac{c}{\sqrt{|c|}}\, \Phi_y}$ is a feasible solution for the matrix $c A$ with objective value $|c|\,W$.  
Also, the block matrices 
$\sfig{\s[\begin{smallmatrix} \Upsilon_x \\ \Upsilon'_x  \end{smallmatrix}]}$ 
and
$\sfig{\s[\begin{smallmatrix} \Phi_y \\ \Phi'_y  \end{smallmatrix}]}$ 
form a feasible solution for the family $A+B$ with objective value at most $W+W'$.

It is obvious that the condition $(\Delta_{xy}=0)\Longrightarrow (A_{xy}=0)$ is necessary for the finiteness of $\gamma_2(A|\Delta)$.  Let us prove its sufficiency.
Assume there is exactly one non-zero matrix in $\{A_{xy}\}$, let it be $A_{ij}$, and, moreover, $A_{ij} = \psi\phi^*$ for some unit vectors $\psi\in\cZ_1$ and $\phi\in\cZ_2$.
We have $\Delta_{ij}\ne 0$.  Let $u$ and $v$ be its left and right singular vectors with maximal singular value, i.e. $u$ and $v$ are unit vectors and $u^*\Delta_{ij}v = \|\Delta_{ij}\|$.  Then 
$
\Upsilon_i = \frac{u\psi^*}{\sqrt{\|\Delta_{ij}\|}}
$
and
$
\Phi_j = \frac{v\phi^*}{\sqrt{\|\Delta_{ij}\|}}
$
is a feasible solution to~\rf(eqn:matrixGamma2) with objective value $1/\|\Delta_{ij}\|$.  From Points~(b) and~(c) of the current proposition, we get the following crude upper bound:
\begin{equation}
\label{eqn:crude}
\gamma_2(A_{xy}|\Delta_{xy}) \le \sum_{x\in\cD_1, y\in\cD_2} \frac{\normtr|A_{xy}|}{\|\Delta_{xy}\|},
\end{equation}
where $0/0=0$.

As a vector norm on a finite-dimensional vector space, $\gamma_2(A|\Delta)$ is continuous in $A$, when restricted to the subspace where it is finite.
By continuity, the triangle inequality is also valid for infinite sums and integrals (as for limits of finite sums).
\pfend

Similarly to the ordinary $\gamma_2$-norm, relative $\gamma_2$-norm can be stated as a semidefinite optimisation problem (SDP).  
As a consequence, it admits an equivalent dual formulation.  We describe it for the case when all $A_{xy}$ are one-dimensional.  Let $A = (a_{xy})$ be a complex $\cD_1\times \cD_2$ matrix, and $\Delta_{xy}$ be as in \rf(defn:matrixGamma2).  
For a complex $\cD_1\times \cD_2$ matrix $\Gamma=(\gamma_{xy})$, let $\Gamma\circ \Delta$ denote the $\cD_1\times \cD_2$ block matrix with blocks $\gamma_{xy}\Delta_{xy}$ (thus, $\Gamma\circ \Delta\colon \bC^{\cD_2}\otimes \cX_2 \to \bC^{\cD_1}\otimes \cX_1$).
The following proposition is proven in \rf(app:Gamma2Scalar):

\begin{prp}
\label{prp:gamma2dual}
In the above notation, $\gamma_2(a_{xy} | \Delta_{xy})_{x\in\cD_1,y\in\cD_2}$ is equal to the optimal value of the following optimisation problem:
\begin{alignat*}{2}
&\mbox{\rm maximise} &\quad&  \norm| \Gamma\circ A |  \\
& \mbox{\rm subject to}&&  \norm| \Gamma\circ \Delta | \le 1,
\end{alignat*}
with the optimisation over $\cD_1\times\cD_2$ complex matrices $\Gamma$.
\end{prp}

Let us end this section with some additional properties of the relative $\gamma_2$-norm, which we use in the paper.

\begin{prp}
\label{prp:gamma2prop}
We have the following properties of the relative $\gamma_2$ norm (omitted subscripts stand for $x\in\cD_1,y\in\cD_2$):
\begin{enumerate}[(a)]
\itemsep0pt
\item \label{stiking@prp:gamma2prop} striking out rows and columns:
$\gamma_2(A_{xy}|\Delta_{xy})_{x\in\cE_1, y\in\cE_2} \le \gamma_2(A_{xy}| \Delta_{xy})$ 
when $\cE_1\subseteq\cD_1$ and $\cE_2\subseteq\cD_2$ ;

\item\label{duplicate@prp:gamma2prop}
duplicating rows and columns:
$\gamma_2(A_{xy}|\Delta_{xy})_{(x,a)\in\cD_1\times\cE_1,\; (y,b)\in\cD_2\times\cE_2} =
\gamma_2(A_{xy}| \Delta_{xy})$;

\item \label{reflexivity@prp:gamma2prop} 
entry-wise lower bound: $\gamma_2(A_{xy} | \Delta_{xy}) \ge \max\limits_{x\in\cD_1,y\in\cD_2} \frac{\|A_{xy}\|}{\|\Delta_{xy}\|}$, where $0/0=0$;\\
in particular, reflexivity: $\gamma_2(A_{xy}|A_{xy}) = 1$ if at least one $A_{xy}$ is non-zero;

\item\label{linear@prp:gamma2prop}
linear transformations:
$\gamma_2(U_x^* A_{xy} V_y\mid \Delta_{xy}) \le \max_{x} \|U_x\| \max_y\|V_y\|\; \gamma_2(A_{xy} |\Delta_{xy})$;

and 
$\gamma_2( A_{xy} | \Delta_{xy}) \le \max_{x} \|U_x\| \max_y\|V_y\|\; \gamma_2(A_{xy} \mid U_x^*\Delta_{xy}V_y)$;

in particular, for any non-zero $c$: $\gamma_2(A_{xy} \mid c\Delta_{xy}) = \frac1{|c|} \gamma_2(A_{xy}|\Delta_{xy})$;

\item \label{composition@prp:gamma2prop} composition property: $\gamma_2(A_{xy}|\Delta_{xy}) \le \gamma_2(A_{xy}|B_{xy})\,\gamma_2(B_{xy}|\Delta_{xy})$;

\item \label{direct@prp:gamma2prop} direct sum property: 
$\gamma_2(A_{xy}\oplus B_{xy}\mid \Delta_{xy}\oplus \mathrm E_{xy}) \le 
\max\sfigA{ \gamma_2(A_{xy}|\Delta_{xy}), \gamma_2(B_{xy} |\mathrm E_{xy}) }$;

in particular, $\gamma_2(A_{xy}\oplus B_{xy}\mid \Delta_{xy}) = 
\max\sfigA{ \gamma_2(A_{xy}|\Delta_{xy}), \gamma_2(B_{xy} |\Delta_{xy}) }$;

\item \label{tensor@prp:gamma2prop} tensor product property:
$\gamma_2(A_{xy}\otimes B_{xy}\mid \Delta_{xy}\otimes \mathrm E_{xy}) \le 
\gamma_2(A_{xy}|\Delta_{xy})\;\gamma_2(B_{xy} |\mathrm E_{xy})$;

\item \label{hadamard@prp:gamma2prop} ``Hadamard product'' property:
$\gamma_2\sA[ {B\elem[x,y]} A_{xy}\mid \Delta_{xy}] \le \gamma_2(B)\; \gamma_2(A_{xy}|\Delta_{xy})$.
\end{enumerate}
\end{prp}

\pfstart
Properties~(\ref{stiking@prp:gamma2prop}) and~(\ref{duplicate@prp:gamma2prop}) are straightforward: One can use a feasible solution to $\gamma_2(A_{xy}| \Delta_{xy})$ for the other problem.

For~(\ref{reflexivity@prp:gamma2prop}), let $\{\Upsilon_x\}$ and $\{\Phi_y\}$ be a feasible solution to $\gamma_2(A_{xy}|\Delta_{xy})$, and $x\in\cD_1$ and $y\in\cD_2$ attain the maximum on the right-hand side of~(\ref{reflexivity@prp:gamma2prop}).  Then, $\|A_{xy}\| \le \|\Upsilon_x\|\|\Delta_{xy}\|\|\Phi_y\|$, hence, at least one of $\|\Upsilon_x\|^2$ and $\|\Phi_y\|^2$ is at least $\frac{\|A_{xy}\|}{\|\Delta_{xy}\|}$.
For the second statement, one can use $\Upsilon_x = \Phi_y =I$ as a feasible solution.

For~(\ref{linear@prp:gamma2prop}), first note that by multiplying all $U_x$ and dividing all $V_y$ by the same positive real number, we may assume that $\max_x \|U_x\|^2 = \max_y \|V_y\|^2 = \max_x \|U_x\|\max_y \|V_y\|$.
Next, let $\{\Upsilon_x\}$ and $\{\Phi_y\}$ be a feasible solution to $\gamma_2(A_{xy}|\Delta_{xy})$.  Then, $\{\Upsilon_x U_x\}$ and $\{\Phi_y V_y\}$ is a feasible solution to $\gamma_2(U_x^* A_{xy} V_y\mid \Delta_{xy})$.
Similarly, if $\{\Upsilon_x\}$ and $\{\Phi_y\}$ is a feasible solution to
$\gamma_2(A_{xy} \mid U_x^*\Delta_{xy}V_y)$, then 
\[
A_{xy} = \Upsilon_x^* ((U_x^* \Delta_{xy} V_y)\otimes I_{\cW}) \Phi_y 
= [(U_x\otimes I_\cW)\Upsilon_x]^*\; (\Delta_{xy}\otimes I_{\cW})\; [(V_y\otimes I_\cW)\Phi_y ].
\]

For~(\ref{composition@prp:gamma2prop}), let $\{\Upsilon_x\}$ and $\{\Phi_y\}$ be a feasible solution to $\gamma_2(A_{xy}|B_{xy})$, and $\{\Upsilon_x'\}$ and $\{\Phi_y'\}$ be a feasible solution to $\gamma_2(B_{xy}|\Delta_{xy})$.  Then, for some $\cW$ and $\cW'$:
\begin{align*}
A_{xy} &= \Upsilon_x^* (B_{xy}\otimes I_{\cW}) \Phi_y 
= \Upsilon_x^* (\Upsilon_x'^*(\Delta_{x,y}\otimes I_{\cW'})\Phi_y'\otimes I_{\cW}) \Phi_y  \\
&= \skA[(\Upsilon_x'\otimes I_{\cW})\Upsilon_x]^* \;(\Delta_{x,y}\otimes I_{\cW\otimes\cW'})\; \skA[(\Phi_y'\otimes I_{\cW})\Phi_y].
\end{align*}

For~(\ref{direct@prp:gamma2prop}) and~(\ref{tensor@prp:gamma2prop}), let $\{\Upsilon_x\}$ and $\{\Phi_y\}$ be a feasible solution to $\gamma_2(A_{xy}|\Delta_{xy})$, and $\{\Upsilon_x'\}$ and $\{\Phi_y'\}$ be a feasible solution to $\gamma_2(B_{xy}|\mathrm E_{xy})$.  We may assume that both solutions use the same space $\cW$.  
Then, 
\[
A_{xy}\oplus B_{xy} = [\Upsilon_x \oplus \Upsilon_x']^*\; ((\Delta_{xy}\oplus \mathrm E_{xy})\otimes I_\cW)\; [\Phi_y \oplus \Phi_y'],
\]
and
\[
A_{xy}\otimes B_{xy} = [\Upsilon_x \otimes \Upsilon_x']^* ((\Delta_{xy}\otimes \mathrm E_{xy})\otimes I_{\cW\otimes \cW}) [\Phi_y \otimes \Phi_y'].
\]

For the second part of~(\ref{direct@prp:gamma2prop}), note that $\gamma_2(A_{xy}\oplus B_{xy}| \Delta_{xy}\oplus \Delta_{xy}) = \gamma_2(A_{xy}\oplus B_{xy}|\Delta_{xy})$, and that
$\gamma_2(A_{xy} | \Delta_{xy}) \le \gamma_2(A_{xy}\oplus B_{xy} | \Delta_{xy})$ 
and
$\gamma_2(B_{xy} | \Delta_{xy}) \le \gamma_2(A_{xy}\oplus B_{xy} | \Delta_{xy})$
due to property~(\ref{linear@prp:gamma2prop}).

Property~(\ref{hadamard@prp:gamma2prop}) is a special case of~(\ref{tensor@prp:gamma2prop}) with one-dimensional matrices $B_{xy} = B\elem[x,y]$ and $\mathrm E_{xy}=1$.
\pfend

\section{General Oracles}
\label{sec:generalOracles}
In this section, we define the adversary bound for problems with general input oracles, which are just arbitrary unitary transformations.  
In \rf(sec:conversion), we study the state conversion problem, and, in \rf(sec:unitary), the unitary implementation problem.
We prove that the adversary bound is semi-tight.
In \rf(sec:unitary), we also derive the tight composition property for the unitary implementation problem.
In \rf(sec:fractional), we consider the fractional query model.

\subsection{State Conversion}
\label{sec:conversion}
In this section, we study the following problem.

\begin{defn}[State Conversion]
\label{defn:conversion}
Let $\cX$ and $\cZ$ be vector spaces, and $\cD$ be some finite set of labels.
For each $x\in\cD$, a unitary operator $O_x$ in $\cX$, and two unit vectors $\rho_x$ and $\sigma_x$ in $\cZ$ are fixed.
\footnote{This notation is borrowed from~\cite{lee:stateConversion}.  Note that $\rho_x$ and $\sigma_x$ are \emph{not} mixed states.}
In the \emph{state conversion} problem $(O_x, \rho_x, \sigma_x)_{x\in\cD}$, the task is to construct a quantum algorithm with an oracle $O$ on $\cX$ satisfying the following constraint.
Assume $\cZ$ is embedded into the state space $\cS\oplus(\cX\otimes\cW)$ of the algorithm in some way (all choices are equivalent).
For each $x\in\cD$, if $O$ is replaced by $O_x$, the algorithm transforms $\rho_x$ into $\sigma_x$.
The \emph{complexity} of the problem is the minimal number of queries used by an algorithm that solves this task.
\end{defn}

We often omit the subindex $x\in\cD$ if it is clear from the context.

Most of the problems we study in this paper are based on this definition by imposing additional constraints on $O_x$, $\rho_x$ or $\sigma_x$.
For instance, the \emph{state generation} problem $(O_x, \sigma_x)_{x\in \cD}$ is defined as the state conversion problem where all $\rho_x = \rho_0$ for some fixed $\rho_0\in\cZ$ (the exact choice of $\rho_0$ is not relevant).

It will be sometimes convenient to relax the condition of $\cD$ being finite.  
The complexity of the problem is then understood as the supremum over all finite subproblems.  The idea is that quantum algorithms are continuous, hence, we can study the problem on a finite $\eps$-net (as an example, see the proof of \rf(thm:unitaryUpper)).

We consider both the \emph{exact} and the \emph{approximate} versions of this problem.  In the latter, some error parameter $\eps>0$ is given, and the algorithm may end its work in some state $\sigma_x'$ satisfying $\|\sigma_x - \sigma'_x\|\le\eps$.
In this section, we develop a bound for this problem with the following \emph{semi-tight} behaviour:
It is a lower bound on the \emph{exact} complexity of the problem and an upper bound on the \emph{approximate} complexity of the problem.

\begin{defn}
\label{defn:advConversion}
For a state conversion problem $(O_x,\rho_x,\sigma_x)_{x\in\cD}$, its \emph{adversary bound} is defined as
\begin{equation}
\label{eqn:advConversion}
\Adv(O_x,\rho_x,\sigma_x)_{x\in\cD} = \gamma_2\sB[ \ip<\rho_x, \rho_y> - \ip<\sigma_x,\sigma_y> \midA O_x - O_y]_{x,y\in\cD}\;.
\end{equation}
\end{defn}

Thus, the adversary bound equals the optimal value of the following optimisation problem:
\begin{subequations}
\label{eqn:advExplicit}
\begin{alignat}{2}
&\mbox{\rm minimise} &\quad& \max \sfigB{ \max\nolimits_{x\in \cD} \norm|u_{x}|^2, \max\nolimits_{y\in\cD} \norm|v_{y}|^2 } \label{eqn:advExplicitObjective} \\
& \mbox{\rm subject to}&&  
\ip<\rho_x, \rho_y> - \ip<\sigma_x,\sigma_y> = \ipB< u_x, \sA[(O_x - O_y)\otimes I_{\cW}] v_y> \qquad \text{\rm for all $x,y\in\cD$;}  \label{eqn:advExplicitCondition} \\
&&& \text{$\cW$ is a vector space,\qquad $u_x, v_y \in \cX\otimes\cW$.}
\end{alignat}
\end{subequations}

\begin{rem}
\label{rem:alternative}
One may replace all $O_x-O_y$ in~\rf(eqn:advConversion) and~\rf(eqn:advExplicitCondition) by $I_\cX-O_x^*O_y$.  This is equivalent as $O_x - O_y = O_x(I_\cX - O_x^*O_y)$, and $O_x$ can be absorbed into $u_x$ (\rfitem(linear@prp:gamma2prop)).
\end{rem}


\begin{thm}
\label{thm:conversionLower}
Assume that a state conversion problem $(O_x,\rho_x,\sigma_x)_{x\in\cD}$ can be solved exactly in $T$ queries.
Then, $\Adv(O_x,\rho_x,\sigma_x)_{x\in\cD} \le T$.
\end{thm}

\pfstart
Assume the vectors $\{\rho_x\}$ are transformed into the vectors $\{\sigma_x\}$ by the algorithm in~\rf(eqn:algorithm).
Its state space is $\cS\oplus(\cX\otimes \cW)$, and $\cZ$ is embedded into it.
For each $x\in\cD$, consider the sequence of quantum states
\begin{equation}
\label{eqn:stateSequence}
\rho^{(0)}_x = \rho_x,\quad \rho^{(1)}_x,\quad \rho^{(2)}_x,\quad\dots\quad \rho^{(T-1)}_x,\quad \rho^{(T)}_x,
\end{equation}
where $\rho_x^{(i)}$ is defined as the state of the quantum algorithm in~\rf(eqn:algorithm) just before the application of $U_i$.
As $\ip<\sigma_x,\sigma_y> = \ipA<\rho^{(T)}_x, \rho^{(T)}_y>$ for all $x,y\in\cD$, 
the problems $\Adv(O_x,\rho_x,\sigma_x)_{x\in\cD}$ and $\Adv\sA[O_x,\rho^{(0)}_x, \rho^{(T)}_x]_{x\in\cD}$ are equal.

Note that the state conversion problem $(O_x, \rho_x^{(i)}, \rho_x^{(i+1)})_{x\in\cD}$ is solved exactly in 1 query.
It suffices to prove the statement of the theorem for each of them, and the general case then follows from the triangle inequality of \rf(prp:gamma2norm) by telescoping.
More precisely, if $\{u_{x}^{(i)}\}$ and $\{v_{y}^{(i)}\}$ is a solution to~\rf(eqn:advExplicit) for the problem $\sA[O_x, \rho_x^{(i)}, \rho_x^{(i+1)}]_{x\in\cD}$, then the solution for the problem $(O_x,\rho_x,\sigma_x)_{x\in\cD}$ is given by
\begin{equation}
\label{eqn:conversionWholeSolution}
u_{x} = \bigoplus\nolimits_{i=0}^{T-1} u_{x}^{(i)}\qquad\text{and}\qquad
v_y = \bigoplus\nolimits_{i=0}^{T-1} v_{y}^{(i)}.
\end{equation}
For each $i$, in dependence on whether the query is direct or reverse,
\[
\text{either}\qquad \rho_x^{(i+1)} = (I_\cS \oplus O_x\otimes I_\cW) U_i \rho_x^{(i)},\qquad\text{or}\qquad\rho_x^{(i+1)} = (I_\cS \oplus O_x^*\otimes I_\cW) U_i\rho_x^{(i)}
\]
holds for all $x\in\cD$.
We consider the first case, the second case being similar. 
We have (where $\Pi$ is the projector onto $\cX\otimes \cW$, and $\Pi^\perp$ is the projector onto $\cS$):
\begin{align*}
\ip<\rho_x^{(i)}, \rho_y^{(i)}> &- \ip<\rho_x^{(i+1)},\rho_y^{(i+1)}> 
=\ipB<U_i\rho_x^{(i)},\, U_i \rho_y^{(i)}> 
- \ipB<(I_\cS \oplus O_x\otimes I_\cW) U_i\rho_x^{(i)},\,(I_\cS \oplus O_y\otimes I_\cW) U_i\rho_y^{(i)}> \\
&=\ipB<\Pi^\perp U_i\rho_x^{(i)},\, \Pi^\perp U_i \rho_y^{(i)}>  + \ipB<(O_x\otimes I_\cW)\Pi U_i\rho_x^{(i)},\, (O_x\otimes I_\cW)\Pi U_i \rho_y^{(i)}> \\
&\qquad -\ipB<\Pi^\perp U_i\rho_x^{(i)},\, \Pi^\perp U_i \rho_y^{(i)}> - \ipB<(O_x\otimes I_\cW) \Pi U_i\rho_x^{(i)},\,(O_y\otimes I_\cW) \Pi U_i\rho_y^{(i)}> \\
&=\ipB<(O_x\otimes I_\cW) \Pi U_i\rho_x^{(i)},\, \sA[(O_x-O_y)\otimes I_\cW]\Pi U_i \rho_y^{(i)}> .
\end{align*}
Thus, we can take
\begin{equation}
\label{eqn:conversionSolution}
u_{x}^{(i)} =  (O_x\otimes I_\cW)\Pi U_i\rho_x^{(i)}\qquad\text{and}\qquad 
v_y^{(i)} = \Pi U_i\rho_y^{(i)}
\end{equation}
as a feasible solution to~\rf(eqn:advExplicit) for the problem $\sA[O_x, \rho_x^{(i)}, \rho_x^{(i+1)}]$.  The norms of these vectors do not exceed 1.
\pfend

To get a lower bound for the approximate version of the problem, one can optimise over all possible target vectors.  For the general case, it is possible to use ideas from~\cite[Section 4.2]{lee:stateConversion}.
We will use this approach in \rf(sec:relations) for more specific problems.

\begin{thm}
\label{thm:conversionUpper}
Let $(O_x, \rho_x, \sigma_x)_{x\in\cD}$ be a state conversion problem with $\Adv(O_x,\rho_x,\sigma_x)<+\infty$, and $0<\eps\le\Adv(O_x,\rho_x,\sigma_x)$ be a real number.
Then, there exists a quantum algorithm with oracle $O$ that, for each $x\in\cD$, if $O$ is replaced by $O_x$, transforms the state $\rho_x$ into a state $\sigma'_x$ satisfying $\norm|\sigma_x -\sigma'_x|\le\eps$.  The algorithm makes $O\s[\Adv(O_x,\rho_x,\sigma_x)\, \eps^{-2}\log\frac1\eps]$ queries.
\end{thm}

As shown by Kothari~\cite{kothari:SDPCharacterization}, the term $\eps^{-2}$ in \rf(thm:conversionUpper) cannot be improved.  
We briefly describe the corresponding construction here.
Consider the following state-generation problem with a standard oracle $O_x$ encoding a string $x\in\{0,1\}^n$ (see \rf(defn:standardOracle)).  The space $\cZ$ is a direct sum of a one-dimensional space $\cA$, and a qubit $\cB$.  The corresponding target state is $\sigma_x = \sqrt{1-1/n}\ket A|0> + \sqrt{1/n}\ket B|\mbox{\scshape{parity}}(x)>$.  It can be shown that $\Adv(O_x, \sigma_x) = O(1)$, while any quantum algorithm that generates $\sigma_x$ with error $\ll\sqrt{1/n}$ has to make $\Omega(n)$ queries.

\pfstart[Proof of \rf(thm:conversionUpper)]
The algorithm is strongly inspired by the algorithm in~\cite{lee:stateConversion}.
Let $\cW$, $\{u_x\}$ and $\{v_y\}$ be a feasible solution to~\rf(eqn:advExplicit) with the objective value $W = \Adv(O_x,\rho_x,\sigma_x)$.
The state space of the algorithm is $\cZ \oplus \cS \oplus \cA \oplus \cB$.
Here the register $\cS$ is isomorphic to $\cZ$, and the registers $\cA$ and $\cB$ are isomorphic to $\cX\otimes \cW$.

Let $\eps' = \eps/4$.
For each $x\in\cD$, define the unit vectors $t_{x+}$ and $t_{x-}$ by 
$t_{x\pm} = \frac1{\sqrt2}\sA[\ket Z|\rho_x> \pm \ket S|\sigma_x>]$,
and the following (non-normalised) vector
\[
\psi_x = t_{x-} + \frac{\sqrt{W/2}}{\eps'} \sA[\ket A |v_x> - \ket B|(O_x\otimes I_\cW) v_x> ] \; .
\]
Let $\Lambda$ be the projector onto the orthogonal complement of the span of the vectors $\{\psi_x\}_{x\in\cD}$.
Let $\Pi_x$ be the projector onto the orthogonal complement of the span of the vectors $\ket A|v> - \ket B|(O_x\otimes I_\cW)v>$ over all $v\in \cX\otimes\cW$.  Let $U_x = (2\Pi_x-I)(2\Lambda-I)$.

Our algorithm executes the phase detection algorithm from \rf(thm:phaseDetection) with the oracle $O = U_x$, precision $\delta = \eps'^2/W$ and error $\eps'$.  After that, the algorithm swaps the register $\cZ$ and $\cS$.  We claim that this algorithm performs the required transformation in the register $\cZ$.

Let us first check the query complexity of the algorithm.  
The transformation $2\Lambda-I$ requires no queries as it is independent of $O$.
The transformation $2\Pi_x-I$ can be implemented in 2 queries.
The query complexity of the algorithm now follows from \rf(thm:phaseDetection).

Now, let us prove the correctness of the algorithm.
Fix an element $x\in\cD$, and, as in \rf(thm:phaseDetection), let $P_{\delta}$ denote the projection onto the eigenvectors of $U_x$ having eigenvalues $\ee^{\ii\theta}$ with $|\theta|\le \delta$.  Denote $P_{\delta}^\perp = I-P_{\delta}$.

\begin{clm}
\label{clm:conversion+}
We have $\| P_0^\perp t_{x+} \| \le \eps'$.
\end{clm}

\pfstart
Consider the vector
\[
\varphi = t_{x+} - \frac{\eps'}{\sqrt{2W}} \sA[\ket A |(O_x^*\otimes I_\cW) u_x> + \ket B|u_x> ].
\]
First, $t_{x+}$ is close to the normalised $\varphi$.  Indeed, using~\rf(eqn:advExplicitObjective), we have
\[
K = \norm|{\frac{\eps'}{\sqrt{2W}} \sA[\ket A |(O_x^*\otimes I_\cW) u_x> + \ket B|u_x> ] }|^2 \le \eps'^2.
\]
Thus,
\(
\ip<t_{x+}, \varphi>^2/\|\varphi\|^2 = 1/(1+K) \ge 1-\eps'^2.
\)
Next, we show that $U_x\varphi = \varphi$, which proves the claim.  For this, note that for all $y\in\cD$,
\[
\ip<\varphi, \psi_y> = \frac12\s[{\ip<\rho_x,\rho_y> - \ip<\sigma_x, \sigma_y> - \ipB<u_x, \sA[(O_x-O_y)\otimes I_\cW] v_y>}] = 0
\]
by~\rf(eqn:advExplicitCondition).
This implies $\Lambda\varphi = \varphi$.  Next, note that
\[
\ket A |(O_x^*\otimes I_\cW) u_x> + \ket B|u_x> = \ket A |w> + \ket B |(O_x\otimes I_\cW)w>
\]
for $w = (O_x^*\otimes I_\cW) u_x$.  Thus, this vector is orthogonal to $\ket A |w> - \ket B |(O_x\otimes I_\cW)w>$, as well as to $\ket A|v> - \ket B|(O_x\otimes I_\cW)v>$ for all $v\perp w$.  Hence, $\Pi_x\varphi = \varphi$.
Together, this proves that $U_x\varphi = \varphi$.
\pfend

\begin{clm}
\label{clm:conversion-}
For all $\delta>0$, we have $\|P_\delta t_{x-} \| \le \frac{\delta}2 \sqrt{1+{W^2}/{\eps'^2}}$.
\end{clm}

\pfstart
This is a simple application of \rf(lem:effective) with $\Pi_1 = \Lambda$, $\Pi_2 = \Pi_x$ and $w = \psi_x$.  
\pfend

Now we can prove the correctness of the algorithm.  Let $\circuit P$ denote the phase detection subroutine used in our algorithm.  Then,
\begin{equation}
\label{eqn:conversionUpperFinal}
\normA| {\circuit P\ket Z|\rho_x> - \ket S|\sigma_x> }| 
= \frac1{\sqrt{2}} \normA| \circuit P(t_{x+} + t_{x-}) - (t_{x+}-t_{x-}) |
\le \frac1{\sqrt{2}}\normA| \circuit Pt_{x+} - t_{x+} | + 
\frac1{\sqrt{2}} \normA| \circuit Pt_{x-} + t_{x-} |.
\end{equation}

By \rf(thm:phaseDetection) and \rf(clm:conversion+), the first term of~\rf(eqn:conversionUpperFinal) is at most $\sqrt{2}\|P_0^\perp t_{x+}\| \le \sqrt{2}\eps'$.  
For the second term, by the same theorem and \rf(clm:conversion-), we have
\[
\frac1{\sqrt{2}} \normA| \circuit Pt_{x-} + t_{x-} |
\le \frac1{\sqrt{2}} \normA| \circuit P P^\perp_\delta t_{x-} + P^\perp_\delta t_{x-} | + 
\sqrt{2} \norm |P_\delta t_{x-}| \le \frac{\eps'}{\sqrt2} + \frac{\delta}{\sqrt{2}} \sqrt{1+\frac{W^2}{\eps'^2}} .
\]
Thus, we have the following estimate on~\rf(eqn:conversionUpperFinal):
\[
\normA| {\circuit P\ket Z|\rho_x> - \ket S|\sigma_x> }|  \le \sqrt{2}\eps' + \frac{\eps'}{\sqrt2} + \frac{\delta}{\sqrt{2}}\sqrt{1+\frac{W^2}{\eps'^2}} \le
\eps'\sB[\sqrt{2}+\frac1{\sqrt2}] + \frac{\eps'^2}{\sqrt{2}W}\sqrt{2\frac{W^2}{\eps'^2}} < 4\eps' = \eps.
 \]
This finishes the proof of \rf(thm:conversionUpper).
\pfend

\subsection{Unitary Implementation and Composition}
\label{sec:unitary}

In this section, we study a variation of the problem from the last section.

\begin{defn}[Unitary Implementation]
\label{defn:unitaryImplementation}
Let $\cX$ and $\cZ$ be vector spaces, and $\cD$ be some finite set of labels.
For each $x\in\cD$, a unitary $O_x$ in $\cX$ and a unitary $V_x$ in $\cZ$ are fixed.
In the \emph{unitary implementation} problem $(O_x, V_x)_{x\in\cD}$, the task is to construct a quantum algorithm with an oracle $O$ on $\cX$ such that, 
for each $x\in\cD$, if $O$ is replaced by $O_x$, the algorithm implements $V_x$ on $\cZ$.
(Again, $\cZ$ is embedded into the space of the algorithm.)
\end{defn}

This problem is essentially a special case of the state conversion problem.
Indeed, we can define the set of labels $\cD'$ as consisting of pairs $(x,\rho)$ with $x\in\cD$ and $\rho$ being a unit vector in $\cZ$.  Then, for a label $(x,\rho)\in\cD'$, we define 
\begin{equation}
\label{eqn:unitaryToConversion}
O_{x,\rho} = O_x,\qquad \rho_{x,\rho} = \rho,\qquad\text{and}\qquad \sigma_{x,\rho} = V_x\rho.
\end{equation}
This state conversion problem is equivalent to the unitary implementation problem.
But as there are many correlations between triples with the same $x$, we can obtain a slightly more compact formulation of the bound.

\begin{defn}
For a unitary implementation problem $(O_x,V_x)_{x\in\cD}$, its \emph{adversary bound} is
\begin{equation}
\label{eqn:advUnitary}
\Adv(O_x,V_x)_{x\in\cD} = \gamma_2\sA[V_x - V_y\mid O_x - O_y]_{x,y\in\cD}\;. 
\end{equation}
\end{defn}

With this definition, we regain the results of \rf(sec:conversion) in the new settings.

\begin{thm}
\label{thm:unitaryLower}
Assume that a unitary implementation problem $(O_x, V_x)_{x\in\cD}$ can be solved exactly in $T$ queries.
Then, $\Adv(O_x,V_x)_{x\in\cD} \le T$.
\end{thm}

\pfstart
Consider the corresponding state conversion problem defined in~\rf(eqn:unitaryToConversion), and the proof of \rf(thm:conversionLower) on this instance.
For a fixed $x$, let $\Upsilon'_x$ map a vector $\rho\in\cZ$ into the vector $u_{x,\rho}$, that is the feasible solution~\rf(eqn:conversionWholeSolution) corresponding to the pair $(x,\rho)\in\cD'$.
Similarly, define $\Phi_y\colon \rho\mapsto v_{y,\rho}$.
From~\rf(eqn:conversionWholeSolution) and~\rf(eqn:conversionSolution), it is apparent that both these transformations are linear.
As $\|u_{x,\rho}\|^2\le T$ for all unit $\rho\in \cZ$, we get that $\|\Upsilon_x'\|^2\le T$.  
Similarly, $\|\Phi_y\|^2\le T$.

By~\rf(eqn:advExplicitCondition), for all $x,y\in\cD$ and unit $\rho, \rho'\in\cZ$, we have
\begin{equation}
\label{eqn:matrixLower1}
\ipB<\Upsilon_x' \rho', \sA[(O_x-O_y)\otimes I_\cW] \Phi_y \rho > = \ip<\rho', \rho> - \ip<V_x\rho', V_y\rho>,
\end{equation}
or, as $\rho$ and $\rho'$ are arbitrary,
\begin{equation}
\label{eqn:matrixLower2}
{\Upsilon'_x}^*\sA[(O_x-O_y)\otimes I_\cW]\Phi_y = I - V_x^*V_y.
\end{equation}
Thus, $\Upsilon_x = \Upsilon'_xV_x^*$ and $\Phi_y$ form a feasible solution to~\rf(eqn:matrixGamma2) with objective value at most $T$.
\pfend

\begin{thm}
\label{thm:unitaryUpper}
Let $(O_x, V_x)$ be a unitary implementation problem with finite $\Adv(O_x,V_x)_{x\in\cD}$, and $\eps>0$ be a real number.
Then, there exists a quantum algorithm with oracle $O$ that, for each $x\in\cD$, if $O$ is replaced by $O_x$, implements an isometry $V_x'$ from $\cZ$ to the state space of the algorithm, such that $\|V_x - V'_x\|\le\eps$.
The algorithm makes $O\s[\Adv(O_x,V_x)\, \eps^{-2}\log\frac1\eps]$ queries.
\end{thm}

\pfstart
This theorem is proven by a direct reduction to \rf(thm:conversionUpper).
Let $\Upsilon_x$ and $\Phi_y$ be the optimal solution to~\rf(eqn:matrixGamma2) for $\Adv(O_x,V_x)$.
For $x\in\cD$ and a unit $\rho\in\cZ$, define vectors
\[
u_{x,\rho} = \Upsilon_x V_x \rho\qquad\text{and}\qquad v_{y,\rho} = \Phi_y \rho.
\]
As in~\rf(eqn:matrixLower1) and~\rf(eqn:matrixLower2), we get that this is a feasible solution to the optimisation problem~\rf(eqn:advExplicit) for the state conversion problem~\rf(eqn:unitaryToConversion).
Also, its objective value is at most $\Adv(O_x,V_x)$.

More precisely, in order to apply \rf(thm:conversionUpper), we need a finite $\cD'$ in the state conversion problem~\rf(eqn:unitaryToConversion).
This is achieved by a standard argument, which we write down explicitly here, and will use implicitly further in the paper.
Select a finite $\eps$-net on $\cZ$ and apply \rf(thm:conversionUpper) for the vectors $\rho$ in the net.  Let $\circuit A$ be the resulting algorithm.
For arbitrary $\rho\in \cZ$, let $\rho'$ be the vector in the net closest to $\rho$.  Then,
\[
\| \circuit A \rho - V_x\rho \| \le
\| \circuit A \rho - \circuit A \rho' \| + \|\circuit A \rho' - V_x\rho' \| + \| V_x \rho' - V_x\rho\| \le 3\eps.
\]
Replacing $\eps$ by $\eps/3$, we obtain the required result.
\pfend

A nice property of~\rf(eqn:advUnitary), compared to \rf(defn:advConversion), is its symmetry of the input and the output.  
It makes possible various applications of the composition property from \rf(prp:gamma2prop).  In particular, the following result is straightforward:

\begin{prp}
\label{prp:composition}
Let $\cD$ be a set of labels, $(O_x, V_x)_{x\in\cD}$ and $(V_x,W_x)_{x\in\cD}$ be unitary implementation problems.  Then,
\(
\Adv(O_x,W_x) \le \Adv(O_x,V_x)\Adv(V_x, W_x).
\)
\end{prp}

A computational interpretation of \rf(prp:composition) is as follows.
Assume we want to implement $(O_x,W_x)$ with constant error. 
Assume we also know solutions to the adversary bounds of $(O_x,V_x)$ and $(V_x, W_x)$ with objective values $T_1$ and $T_2$, respectively.
One approach is to use \rf(thm:unitaryUpper) to get an $O(T_2)$ query algorithm $\circuit B$ for the problem $(V_x,W_x)$.
Then, we can use \rf(thm:unitaryUpper) again to get an algorithm $\circuit A$ for the problem $(O_x,V_x)$ and use it as an oracle for $\circuit B$.  In order for this to work, we must set the error of the algorithm $\circuit A$ to $\eps \ll 1/T_2$.  Thus, the complexity of one execution of $\circuit A$ becomes $\tO(T_1T_2^2)$, and the total complexity of the algorithm becomes $\tO(T_1T_2^3)$.  

\rf(prp:composition) shows that we can actually get an algorithm with complexity $O(T_1T_2)$ instead, thus saving a $T_2^2$ factor!
This is more impressive than the logarithmic savings for function evaluation.  Moreover, if we only have an approximate algorithm for the problem $(O_x, V_x)$, and don't know a solution to the adversary bound, we, in general, cannot amplify the precision of the algorithm, and cannot compose the two procedure at all.

It is possible to get other results in the spirit of \rf(prp:composition).
The next proposition corresponds to the observation that the query complexity of implementing two subroutines in superposition equals the complexity of the hardest of them, and that implementing a tensor product of them is at most the sum of the complexities of the two.
\begin{prp}
Let $\cD$ be a set of labels, $(O_x, V_x)_{x\in\cD}$ and $(O_x,W_x)_{x\in\cD}$ be unitary implementation problems.  Then,
\[
\Adv(O_x,V_x\oplus W_x) = \max\sfigA{\Adv(O_x,V_x), \Adv(O_x, W_x)}
\]
and
\[
\Adv(O_x,V_x\otimes W_x) \le \Adv(O_x,V_x) + \Adv(O_x, W_x).
\]
\end{prp}

\pfstart
The first equation immediately follows from \rfitem(direct@prp:gamma2prop).
For the second inequality, using Propositions~\rfitemE(triangle@prp:gamma2norm) and~\rfitemE(tensor@prp:gamma2prop), we have (all relative $\gamma_2$-norms are on $x,y\in\cD$):
\begin{align*}
\Adv(O_x,V_x\otimes W_x) &=\gamma_2\sA[V_x\otimes W_x - V_y\otimes W_y \mid O_x - O_y]\\
&\le \gamma_2\sA[V_x\otimes W_x - V_y\otimes W_x \mid O_x - O_y] +
\gamma_2\sA[V_y\otimes W_x - V_y\otimes W_y \mid O_x - O_y] \\
&\le \gamma_2\sA[V_x - V_y \mid O_x - O_y]\gamma_2(W_x\mid 1) +
\gamma_2\sA[W_x - W_y \mid O_x - O_y]\gamma_2(V_y\mid 1) \\
&\le \Adv(O_x,V_x) + \Adv(O_x, W_x),
\end{align*}
since from $\|W_x\|=1$ we get $\gamma_2(W_x\mid 1)_{x,y\in\cD} =1$, and similarly for $\gamma_2(V_y\mid 1)$.
\pfend

With this machinery, it is also easy to introduce costs of various subroutines in a way similar to~\cite{reichardt:spanPrograms}:

\begin{prp}
Let $\cD$ be a set of labels, $(O_x, V^{(i)}_x)_{x\in\cD}$, for $i\in[n]$, and $(O_x, W_x)_{x\in\cD}$ be unitary implementation problems.  Then,
\[
\Adv(O_x, W_x) \le \gamma_2\sC[ W_x - W_y \midB \bigoplus_{i\in[n]} \frac{V^{(i)}_x - V^{(i)}_y}{\Adv\sA[O_x, V^{(i)}_x]} ]_{x,y\in\cD}\;.
\]
\end{prp}

Intuitively, the last expression means that it costs $\Adv\sA[O_x, V^{(i)}_x]$ queries to execute the subroutine that implements $V_x^{(i)}$.

\pfstart
By Propositions~\rfitemE(homogen@prp:gamma2norm) and~\rfitemE(direct@prp:gamma2prop), we have
\[
\gamma_2\sC[ \bigoplus_{i\in[n]} \frac{V^{(i)}_x - V^{(i)}_y}{\Adv\sA[O_x, V^{(i)}_x]} \midB O_x-O_y]_{x,y\in\cD} = 1,
\]
and the statement follows from~\rfitem(composition@prp:gamma2prop).
\pfend

\subsection{Fractional Query Model}
\label{sec:fractional}
In this section, we briefly discuss how the results of the previous sections can be applied for the \emph{fractional} query model.
The fractional query model is equivalent to the more known \emph{continuous} query model for the standard input oracle~\cite{cleve:discreteTimeSimulation, berry:exponentialSimulatingHamiltonians}.  A similar equivalence still holds for general input oracles.
Let us consider the unitary implementation problem, the state conversion problem being similar.

Let $\cD$ be a set of labels.  For $x\in\cD$, we have to implement a unitary $V_x$.  This time, instead of the input oracle $O_x$, we have an input Hamiltonian $H_x$, which is an arbitrary Hermitian operator.  A query algorithm alternates unitary transformations and executions of the input Hamiltonian.  The algorithm can execute the input Hamiltonian for arbitrary time $t>0$, which results in the application of the unitary $\ee^{-\ii Ht}$.  The complexity of the algorithm is the total execution time of the input Hamiltonian.  

We may assume that all applications of the input Hamiltonian have some small duration $t$.  The adversary bound of this problem is then equal to
\[
\lim_{t\to +0} \; t \,\gamma_2\sB[V_x - V_y \midA \ee^{-\ii H_x t} - \ee^{-\ii H_y t}]_{x,y\in\cD} =
\lim_{t\to +0} \gamma_2\sB[V_x - V_y \midB \frac{\ee^{-\ii H_x t} - I_\cX}{t} - \frac{\ee^{-\ii H_y t} - I_{\cX}}t]_{x,y\in\cD},
\]
where we used \rfitem(linear@prp:gamma2prop).  Clearly,
\[
\lim_{t\to +0} \frac{\ee^{-\ii H_x t} - I_\cX}{t} = \frac{\mathrm d}{\mathrm dt}\, \ee^{-\ii H_x t}\midA_{t=0} = -\ii H_x.
\]
Using \rf(lem:gamma2continuous) below and \rfitem(homogen@prp:gamma2norm), the adversary bound of implementation of $\{V_x\}$ in the fractional query model with input Hamiltonians $\{H_x\}$ is given by
\[
\gamma_2\sA[ V_x - V_y \mid H_x - H_y ]_{x,y\in\cD}\;.
\]


\begin{lem}
\label{lem:gamma2continuous}
Let $\cD_1$ and $\cD_2$ be finite sets of labels, and $R\subseteq \cD_1\times \cD_2$.
Consider the subspace $\cM$ of families of matrices $\Delta = \{\Delta_{xy}\}_{x\in\cD_1,y\in\cD_2}$ satisfying $\Delta_{xy}=0$ for all $(x,y)\in R$.
Let $\cM_0$ denote the subset of all $\{\Delta_{xy}\}\in\cM$ satisfying $\Delta_{xy}\ne 0$ for all $(x,y)\notin R$.
For a fixed family $A = \{A_{xy}\}_{x\in\cD_1,y\in\cD_2}$, the function $\Delta\mapsto \gamma_2(A|\Delta)$ of $\Delta$ on $\cM$, is continuous in all points of $\cM_0$.
\end{lem}

\pfstart
If $A_{xy}\ne 0$ for some $(x,y)\in R$, then $\gamma_2(A|\Delta)=+\infty$ on all $\cM$, hence, it is continuous.  Let us now assume that $A_{xy} = 0$ for all $(x,y)\in R$.

Fix a parameter $\delta>0$, and denote by $\cM_\delta$ the subset of all $\{\Delta_{xy}\}\in\cM$ satisfying $\|\Delta_{xy}\|>\delta$ for all $(x,y)\notin R$.
We will show that $\gamma_2(A|\Delta)$ is uniformly continuous on $\cM_\delta$.  As $\delta$ is arbitrary, this proves that $\gamma_2(A|\Delta)$ is continuous on $\cM_0$.

Let $\Delta = \{\Delta_{xy}\}$ and $\Delta' = \{\Delta'_{xy}\}$ be two families in $\cM_\delta$.  
Let $\{\Upsilon_x\}$ and $\{\Phi_y\}$ be an optimal solution to $\gamma_2(A|\Delta)$.  Then, by \rfitem(triangle@prp:gamma2norm), where the relative $\gamma_2$-norms are with respect to $x\in\cD_1, y\in\cD_2$,
\begin{align*}
\gamma_2(A|\Delta') & \le \gamma_2 \sB[ \Upsilon_x^* (\Delta'_{xy}\otimes I_{\cW}) \Phi_y \midA \Delta'_{xy} ] + \gamma_2 \sB[ \Upsilon_x^* {\sA[(\Delta_{xy} - \Delta'_{xy})\otimes I_{\cW}]} \Phi_y \midA \Delta'_{xy} ]\\
&\le \gamma_2(A|\Delta) + 
\gamma_2 \sB[ \Upsilon_x^* {\sA[(\Delta_{xy} - \Delta'_{xy})\otimes I_{\cW}]} \Phi_y \midA \Delta'_{xy} ] .
\end{align*}
Applying~\rf(eqn:crude) twice, we see that
\begin{align*} 
\gamma_2 \Bigr( \Upsilon_x^* {\sA[(\Delta_{xy} - \Delta'_{xy}) \otimes I_{\cW}]} \Phi_y \midA \Delta'_{xy}\Bigl) &\le \frac{|\cD_1|\cdot|\cD_2|}{\delta} \max_{x,y} \normtrB| \Upsilon_x^* {\sA[(\Delta_{xy} - \Delta'_{xy})\otimes I_{\cW}]} \Phi_y | \\
&\le \frac{|\cD_1|\cdot|\cD_2|}{\delta} \max_{x,y} \s[\dim \cZ_1 \|\Upsilon_x^*\|\; \|\Phi_y\|\; \|\Delta_{xy} - \Delta'_{xy}\|  ] \\
&\le \frac{|\cD_1|\cdot|\cD_2| \sum_{xy} \normtr|A_{xy}|} {\delta^2} \dim \cZ_1 \max_{x,y}\|\Delta_{xy} - \Delta'_{xy} \|.
\end{align*}
As we can exchange $\Delta$ and $\Delta'$, $\gamma_2(A|\Delta)$ is uniformly continuous on $\cM_\delta$.
\pfend

Let us note that, in general, $\gamma_2(A|\Delta)$ is \emph{not} continuous in $\Delta$ outside $\cM_0$.
Indeed, let $A=(a_{xy})$ be a $01$-matrix.
By \rfitem(reflexivity@prp:gamma2prop), $\gamma_2\sA[a_{xy}|a_{xy}] = 1$.
Now let $\eps>0$ be a small real number, and $\Delta_{xy} = (\delta_{xy})$ be defined by $\delta_{xy}=1$ if $a_{xy}=1$, and $\delta_{xy}=\eps$ if $a_{xy}=0$.  In this case, using~\rf(eqn:1dgamma2),  
$\gamma_2\sA[a_{xy}|\delta_{xy}] = \gamma_2(A)$.
Finally, it is possible to construct a $01$-matrix with arbitrary large $\gamma_2$-norm, for example, $\gamma_2\s[\begin{smallmatrix} 1 & 0 \\ 1 & 1\end{smallmatrix}]>1$.

\section{State-Generating Oracles}
\label{sec:stateOracles}

In this section, we study a more specific problem, in which we are given an oracle that prepares some quantum state, or a direct sum of a number of such oracles.
An important special case is the standard input oracle that encodes a sting $x\in [q]^n$.
This oracle is used by most of quantum algorithms and previous research on quantum adversaries, starting with the hybrid method and ending with~\cite{lee:stateConversion}, focused on this case.

We proceed by applying general results of \rf(sec:generalOracles) to this special case.
We get two different formulations of the adversary bound for this problem.
The first formulation is easier to use, and it has composability properties similar to \rf(prp:composition), but it can be much larger than the true adversary bound.
Another one is a tight characterisation of the adversary bound, but it is also harder to use.

In \rf(sec:counterexample), we compare performance of both bounds on a simple problem, and show that, although the bounds we got in Theorems~\ref{thm:conversionUpper} and~\ref{thm:unitaryUpper} are similar, the bound for unitary implementation is more precise for some problems.



\subsection{Definitions and Bounds}
\label{sec:preparation}
In this section, we will work with the following input oracle:

\begin{defn}[State-Generating Oracle]
\label{defn:statePreparation}
Let $\cX$ be a vector space with a fixed unit vector $e_0\in\cX$.
An oracle $O_x$ \emph{generates} a state $\psi_x$ iff $O_xe_0 = \psi_x$.

We will also consider direct sums of such oracles.
That is, let $\cJ = \bC^{n}$, and $O_x$ acting on $\cJ\otimes\cX$ is decomposable into $O_x = \bigoplus_{j\in [n]} O_{x,j}$, where each $O_{x,j}$ acts on $\cX$ and generates a state $\psi_{x,j}$.
In particular, we have $O_x\ket J|j>\ket X|0> = \ket J|j>\ket X|\psi_{x,j}>$ for all $j\in[n]$.
\end{defn}

The convention is that the algorithm works for any oracle generating the states $\psi_{x,j}$.
The case $n=1$ is the fundamental one, and the general case can be reduced to it by \rfitem(direct@prp:gamma2prop).
We will assume that $e_0$ is orthogonal to all $\psi_{x,j}$.  This is without loss of generality, as one can always add one more dimension to $\cX$.
The most common variant of the state-generating oracle is as follows.

\begin{defn}[Oracle Encoding a String]
\label{defn:standardOracle}
Let $\cX$ has orthonormal basis $e_0,\dots,e_q$.
A \emph{standard} oracle $O_x$ \emph{encoding} a string $x = (x_j)\in[q]^n$ is a state-generating oracle of \rf(defn:statePreparation) with $\psi_{x,j} = e_{x_j}$.

Let $\cH$ be an arbitrary vector space.
A \emph{noisy} oracle $\tO_x$ encoding a string $x\in[q]^n$ is a state-generating oracle with each $\psi_{x,j}\in\cX\otimes\cH$ satisfying $\|(\Pi_{x_j}\otimes I_{\cH})\psi_{x,j}\|^2\ge \frac12+\Omega(1)$.  Here $\Pi_a$ is the projector $e_a e_a^*$ on $\cX$.
\end{defn}

If the input oracle $O_x$ of a state conversion or a unitary implementation problem is a state-generating oracle, we replace it by the list of the corresponding vectors, i.e., we write $\sA[\{\psi_{x,j}\}_{j\in[n]}, \rho_x, \sigma_x]_{x\in\cD}$ or $\sA[\{\psi_{x,j}\}_{j\in[n]}, V_x]_{x\in\cD}$.
We omit the subscripts if they are clear from the context.
For instance, for the standard input oracle, we get the problem $\sA[\{e_{x_j}\}, \rho_x, \sigma_x]$ studied in~\cite{lee:stateConversion}.
Our task in this section is to provide estimates on the adversary bound with state-generating oracles.


\begin{prp}
\label{prp:preparationEasy}
For a state conversion 
$\sA[\{\psi_{x,j}\}_{j\in[n]}, \rho_x, \sigma_x]_{x\in\cD}$
and a unitary implementation
$\sA[\{\psi_{x,j}\}_{j\in[n]}, V_x]_{x\in\cD}$
problems, we have the following estimates:
\begin{align}
\Adv\sA[\{\psi_{x,j}\}, \rho_x, \sigma_x] 
&\le \gamma_2\s[ \ip<\rho_x,\rho_y>-\ip<\sigma_x, \sigma_y> \midB \bigoplus\nolimits_{j\in[n]} {\sA[1-\ip<\psi_{x,j},\psi_{y,j}>]}]_{x,y\in\cD}
\label{eqn:preparationEasy}\\
\Adv\sA[\{\psi_{x,j}\}, V_x] 
&\le \gamma_2\s[ V_x-V_y \midB \bigoplus\nolimits_{j\in[n]} {\sA[1-\ip<\psi_{x,j},\psi_{y,j}>]}]_{x,y\in\cD}.
\label{eqn:preparationUnitaryEasy}
\end{align}
\end{prp}

The direct sum in~\rf(eqn:preparationEasy) and~\rf(eqn:preparationUnitaryEasy) stands for an $n\times n$ diagonal matrix.

\pfstart
Let $\cD'$ denote the label set of the state conversion problem corresponding to $\sA[\{\psi_{x,j}\}, \sigma_x]$, i.e., each label consists of $x\in\cD$ and the choice of a unitary $O_x = \bigoplus_{j\in[n]} O_{x,j}$ preparing the vectors $\psi_{x,j}$.
By \rf(thm:conversionUpper), for each $j\in[n]$, we have
$\gamma_2\sA[1-\ip<\psi_{x,j},\psi_{y,j}>\mid O_{x,j}-O_{y,j}]_{x,y\in\cD'} \le 1$.
By \rfitem(direct@prp:gamma2prop),
\[
\gamma_2\s[\bigoplus\nolimits_{j\in[n]} (1-\ip<\psi_{x,j},\psi_{y,j}>) \midB O_x - O_y]_{x,y\in\cD'} \le 1.
\]
The statement now follows from Points~(\ref{duplicate@prp:gamma2prop}) and (\ref{composition@prp:gamma2prop}) of \rf(prp:gamma2prop).
\pfend

\mycommand{tAdv}{\mathop{\widetilde{\mathrm{Adv}}}}

Inspired by this result, let us denote by 
$\tAdv\sA[\{\psi_{x,j}\}, \rho_x, \sigma_x]$ 
and
$\tAdv\sA[\{\psi_{x,j}\}, V_x]$ 
the right-hand sides of~\rf(eqn:preparationEasy) and~\rf(eqn:preparationUnitaryEasy), respectively.
An equivalent formulation of $\tAdv\sA[\{\psi_{x,j}\}, \rho_x, \sigma_x]$, which is closer in form to the usual formulation of the dual adversary bound, is
\begin{subequations}
\label{eqn:advOrig}
\begin{alignat}{3}
&\mbox{\rm minimise} &\quad& \max_{x\in \cD} \max\sfigB{ \sum\nolimits_{j\in [n]}\|u_{x,j}\|^2, \sum\nolimits_{j\in [n]}\|v_{x,j}\|^2 } \label{eqn:advOrigObjective} \\
& \mbox{\rm subject to}&& \sum\nolimits_{j\in[n] } \ip<u_{x,j}, v_{y,j}> \sA[1-\ip<\psi_{x,j},\psi_{y,j}>] = \ip<\rho_x,\rho_y> - \ip<\sigma_x,\sigma_y> &\quad& \text{\rm for all $x,y\in\cD$.} \label{eqn:advOrigCondition}\\
&&& \text{$\cW$ is a vector space,\qquad $u_{x,j}, v_{y,j} \in \cW$.} 
\end{alignat}
\end{subequations}
This bound has some nice properties.  It is easier than the general definition of $\Adv\sA[\{\psi_{x,j}\}, \rho_x, \sigma_x]$.  
For a state generation problem, it has a symmetry between the input and the output, like the expression in~\rf(eqn:advUnitary).  In particular, it can be composed with itself akin \rf(prp:composition).

Inspired by this symmetry, one might guess that this is a tight bound on $\Adv\sA[\{\psi_{x,j}\}, \rho_x, \sigma_x]$.
But it turns out that it is \emph{not} tight, as we will show in \rf(sec:counterexample).
On the other hand, there is one important special case, when it is.

\begin{prp}
\label{prp:preparationOrhtogonal}
Assume that $\cD\subseteq[q]^n$, and a state conversion problem $\sA[\{e_{x_j}\}_{j\in[n]}, \rho_x, \sigma_x]_{x\in\cD}$ with the standard oracle is given.  
Then, $\Adv\sA[\{e_{x_j}\}, \rho_x, \sigma_x] \ge \frac12 \tAdv\sA[\{e_{x_j}\}, \rho_x, \sigma_x]$, that is,
\[
\Adv\sA[\{e_{x_j}\}, \rho_x, \sigma_x] \ge \frac12\; \gamma_2\s[ \ip<\rho_x,\rho_y>-\ip<\sigma_x, \sigma_y> \midB \bigoplus\nolimits_{j\in[n]} 1_{x_j\ne y_j}]_{x,y\in\cD}.
\]
A similar result holds for the unitary implementation problem.
\end{prp}

Thus, we obtain the formulation of the adversary bound from~\cite{lee:stateConversion}.

\pfstart
Similarly to the proof of \rf(prp:preparationEasy), it suffices to show that
\[
\gamma_2\sA[ O_{x,j}-O_{y,j} \mid 1_{x_j\ne y_j}]_{x,y\in\cD'} \le 2.
\]
To show this, note that the block matrices 
$\Upsilon_x = \s[\begin{smallmatrix} O_{x,j}^* \\ I_{\cX}  \end{smallmatrix}]$ 
and
$\Phi_y = \s[\begin{smallmatrix} I_{\cX} \\ -O_{y,j}  \end{smallmatrix}]$ 
form a feasible solution to~\rf(eqn:matrixGamma2) in this case.
\pfend

There is an alternative way to estimate $\Adv(\{\psi_{x,j}\}, \sigma_x)$.  For $\psi\in\cX$, orthogonal to $e_0$, let $R(\psi) = I_\cX - (e_0-\psi)(e_0-\psi)^*$ be the reflection through the orthogonal complement of $(e_0-\psi)/\sqrt{2}$.

\begin{prp}
\label{prp:preparationHard}
For a unitary implementation problem $\sA[\{\psi_{x,j}\}_{j\in[n]}, V_x]_{x\in\cD}$, we have
\[
\frac12\Adv\sA[\{\psi_{x,j}\}, V_x] 
\le \gamma_2\sC[ V_x - V_y \midB \bigoplus_{j\in[n]} {\sA[R(\psi_{x,j}) - R(\psi_{y,j})]}]_{x,y\in\cD}
\le \Adv\sA[\{\psi_{x,j}\}, V_x].
\]
\end{prp}
  
\pfstart
The first inequality, similarly to \rf(prp:preparationEasy), is a consequence of
\[ 
\gamma_2\sA[ R(\psi_{x,j}) - R(\psi_{y,j}) \mid O_{x,j}-O_{y,j}]_{x,y\in\cD'}\le 2.
\]
The latter follows from \rf(thm:unitaryLower), as given an oracle $O_{x,j}$ mapping $e_0$ into $\psi_{x,j}$, one can reflect about $(e_0-\psi_{x,j})/\sqrt{2}$ in two queries.  (Recall that we assumed that $\psi_{x,j}$ is orthogonal to $e_0$.)

For the second inequality, note that $R(\psi_{x,j})$ transforms $e_0$ into $\psi_{x,j}$, and, thus, is a valid $O_{x,j}$ from \rf(defn:statePreparation).
\pfend

\subsection{Comparison of The Bounds}
\label{sec:counterexample}

In this section, we analyse the performance of bounds from \rf(sec:preparation) on the following simple problem.
Let $n=1$, and consider the following state generation problem $(\psi_x, \sigma_x)_{x\in\cD}$,
where we are given an oracle that generates a state $\psi_x$, and the task is to generate $\sigma_x$.
For simplicity, we assume all vectors have real entries.

The problem is parametrised by a positive real number $\alpha$, that we assume to be small.
The set of labels $\cD$ has a special label 0.
For all $x\in\cD\setminus\{0\}$, we have $\ip<\psi_x, \psi_0>=\cos\alpha$.
The target states satisfy $\sigma_x = \sigma_y$ and $\sigma_x \perp \sigma_0$ for all $x,y\ne 0$.
For simplicity, we assume that $\alpha = \pi/(4k)$ for a positive integer $k$.

\begin{clm}
The above problem can be solved exactly in $O(1/\alpha)$ queries.
\end{clm}

\pfstart
This is an application of exact quantum amplitude amplification~\cite{brassard:amplification}.
For completeness, let us describe the algorithm.
Start in the state $\psi_0$ and repeatedly reflect about $\psi_0$ and $\psi_x$.
If $x=0$, the state $\psi_0$ remains intact.
If $x\ne0$, each iteration of two reflections performs a rotation by angle $2\alpha$ in the plane spanned by $\psi_0$ and $\psi_x$.  Thus, after $k$ iterations, the state $\phi$ of the algorithm is orthogonal to $\psi_0$.  Attach a fresh qubit containing $0$ if $\phi=\psi_0$ and $1$ if $\phi\perp \psi_0$, and run the whole procedure in reverse.
\pfend

\begin{clm}
\label{clm:tAdvNonTight}
For the above problem, we have $\tAdv(\psi_x, \sigma_x) = \Theta(1/\alpha^2)$.
\end{clm}

\pfstart
Let $x$ be an arbitrary non-zero element of $\cD$.
Note that $1-\ip<\psi_x,\psi_0> = 1-\cos\alpha \approx \alpha^2$.
On the other hand, $1-\ip<\sigma_x,\sigma_0> = 1$.
Hence, by \rfitem(reflexivity@prp:gamma2prop), $\tAdv(\psi_x,\sigma_x)=\Omega(1/\alpha^2)$.

To get the upper bound, use the following matrices as a feasible solution
\[
\Upsilon_0 = \Phi_x = \begin{pmatrix} 1/\sqrt{1-\cos\alpha} \\ 0  \end{pmatrix}
\qquad\text{and}\qquad
\Phi_0 = \Upsilon_x = \begin{pmatrix} 0 \\ 1/\sqrt{1-\cos\alpha}  \end{pmatrix}
\]
for all $x\ne 0$.
\pfend

Thus, \rf(clm:tAdvNonTight) shows that the estimate in \rf(prp:preparationEasy) is not tight.
On, the other hand, we know that the estimate from \rf(prp:preparationHard) is tight, and we will show an explicit solution.
For this, the following easy lemma is useful.  We omit the proof.

\begin{lem}
\label{lem:reflectionDifference}
Let $R_\psi$ and $R_\phi$ be reflections about the orthogonal complements of vectors $\psi$ and $\phi$ in $\cX$.  Let $\cS\subseteq \cX$ be the span of $\psi$ and $\phi$, and let $\alpha$ be the angle between them.
The difference $R_\psi-R_\phi$ is zero on $\cS^\perp$, and it has eigenvalues $\pm2\sin\alpha$ on $\cS$ with eigenvectors that are the bisectors of $\psi$ and $\phi^\perp$, the latter complement taken in $\cS$.
\end{lem}

\begin{clm}
\label{clm:RxAdvIsTight}
For $x\in\cD$, let us denote $R_x = R(\psi_x)$ the reflection through the orthogonal complement of $\psi'_x = (e_0 - \psi_x)/\sqrt{2}$.
Then, $\gamma_2(1-\ip<\sigma_x,\sigma_y> \mid R_x - R_y)_{x,y\in\cD} = \Theta(1/\alpha)$.
\end{clm}

\pfstart
Note that, for $x\ne0$, the angle between $\psi'_x$ and $\psi'_0$ is $\beta\approx \alpha/\sqrt{2}$.
By \rf(lem:reflectionDifference), $R_0-R_x$ has eigenvalues $\pm 2\sin\beta$ with the eigenvectors being the bisectors of ${\psi'_0}^\perp$ and $\psi'_x$.  
From \rfitem(reflexivity@prp:gamma2prop), we get a lower bound $\gamma_2\sA[1-\ip<\sigma_x,\sigma_y> \mid R_x - R_y] \ge 1/(2\sin\beta) = \Omega(1/\alpha)$.

Let $\varphi_x$ be the eigenvector of $R_0-R_x$ with eigenvalue $2\sin\beta$.  If $\alpha$ is small, the angle between $\varphi_x$ and $\psi'_0$ is around $\pi/4$.
Note that $c = \ipA<\varphi_x, (R_0 - R_x) \psi'_0>$ is independent of $x$, and 
$c \approx \sqrt{2}\sin\beta\approx \alpha$.
Thus, we can take the following feasible solution
\[
\Upsilon_0 = \begin{pmatrix} \psi'_0/\sqrt{c} \\ 0  \end{pmatrix},\quad
\Phi_x = \begin{pmatrix} \varphi_x/\sqrt{c} \\ 0  \end{pmatrix},\quad
\Phi_0 = \begin{pmatrix} 0\\ \psi'_0/\sqrt{c}   \end{pmatrix},\quad\text{and}\quad
\Upsilon_x = \begin{pmatrix} 0\\ \varphi_x/\sqrt{c}   \end{pmatrix}
\]
to show that $\gamma_2\sA[1-\ip<\sigma_x,\sigma_y> \mid R_x - R_y] = O(1/\alpha)$.
(Here we used that $R_x - R_y$ is self-adjoint.)
\pfend

This proves that \rf(prp:preparationHard) is more precise than \rf(prp:preparationEasy) for vectors that are close to each other.
Essentially, this is because $\|R_x-R_0\|$ scales as $\alpha$, whereas $1-\ip<\psi_x,\psi_0>$ scales as $\alpha^2$.
On the other hand, since the location of the eigenvectors of $R_x-R_y$ depends on both $x$ and $y$, it is hard to ``catch'' them by a solution $\{\Upsilon_x\}$, $\{\Phi_y\}$ with parts that depend on $x$ and $y$ separetely.

This has other consequences.
This time, consider a different state generation problem.
Let $\cD = \{0,1\}$, $\ip<\psi_0,\psi_1> = \cos\alpha$, and $\ip<\sigma_0,\sigma_1>=\cos\gamma$, where $\gamma$ satisfies $1-\cos\gamma = \sin\alpha$, that is, $\gamma\approx\sqrt{\alpha}$.
As in \rf(clm:RxAdvIsTight), we get that
\begin{equation}
\label{eqn:counter1}
\gamma_2\sA[1-\ip<\sigma_x,\sigma_y> \mid R_x - R_y]_{x,y\in\cD}=\Theta(1).
\end{equation}
On the other hand, by the proof of the same claim, for any unitaries $V_0$ and $V_1$ that transform $\rho_0$ into $\sigma_0$ and $\sigma_1$, respectively, we have that 
\begin{equation}
\label{eqn:counter2}
\gamma_2\sA[V_x - V_y \mid R_x - R_y]_{x,y\in\cD}=\Theta(1/\sqrt{\alpha}).
\end{equation}
Thus, comparing~\rf(eqn:counter1) and~\rf(eqn:counter2), we arrive at a paradoxical conclusion:
The adversary bound allows generation of $\sigma_0$ and $\sigma_1$ in a constant number of queries, but simultaneously prohibits any unitary that would perform this task!
Or, to put it less dramatically, \rf(thm:unitaryLower) is more precise than \rf(thm:conversionLower).

Also, in sight of \rf(eqn:counter2) and \rf(thm:unitaryLower), an algorithm (like the one in \rf(thm:conversionUpper)) that performs the transformation in~\rf(eqn:counter1) with error $\ll \sqrt{\alpha}$ has to make $\Omega(1/\sqrt{\alpha})$ queries.  
This shows that the $\eps^{-2}$ term in the complexity estimate of \rf(thm:conversionUpper) cannot be improved to less than $\eps^{-1}$.

\section{Evaluation of Functions and Relations}
\label{sec:relations}
In this section, we study the problem of evaluating relations and functions with emphasis on  lower bounds.
In \rf(sec:relationBasic), we obtain lower bounds by optimising the adversary bound for state generation over the set of allowed target states and taking the dual.
Our approach in \rf(sec:purifiers) is more interesting.
We show that the adversary bound is a lower bound not only for the exact version of the problem, but also for the approximate one in two cases: When we evaluate a function; or when we evaluate a relation, provided that we can effectively test the output of the algorithm.
This is done by constructing the adversary bound for the problem of outputting the exact solution, given an oracle that prepares an approximate solution.

\subsection{Definitions and Basic Bounds}
\label{sec:relationBasic}

\begin{defn}[Evaluation of Relation]
\label{defn:relation}
Let $\cX$ be a vector space, $\cD$ be a set of labels, and $\{O_x\}_{x\in\cD}$ be a collection of input oracles on $\cX$.
Assume also that there is given a relation $r\colon \cD\to [m]$.
A quantum algorithm that \emph{evaluates} the relation $r$ with input oracles $\{O_x\}$ is any state-generating algorithm from \rf(defn:conversion) with $\cZ = \cY\otimes\cH$, where $\cY = \bC^m$ and $\cH$ is arbitrary, that satisfies the following condition.
The state $\rho_x$ equals some fixed $\rho_0$ for all $x\in\cD$, and $(\Pi_{r(x)}\otimes I_{\cH}) \sigma_x = \sigma_x$ for all $x\in\cD$.  Here $\Pi_{r(x)}$ is the projector $\sum_{a\in r(x)} e_ae_a^*$ in $\cY$.

Evaluation of a function is a special case of this definition with $|r(x)|=1$ for all $x\in\cD$.
\end{defn}

We say that the above algorithm has error $\eps$ if $\| (\Pi_{r(x)}\otimes I_{\cH})\sigma_x\|^2 \ge 1-\eps$ for all $x\in\cD$.  Note that this notion is different than the one assumed in Sections~\ref{sec:generalOracles} and~\ref{sec:stateOracles}, but it is conventional in quantum algorithms.

Throughout the section, we use notation $\Delta=(\Delta_{xy})_{x,y\in\cD}$ for the collection of matrices $\Delta_{xy} = O_x - O_y$.
Using \rf(defn:relation) and \rf(thm:conversionLower), it is straightforward to lower bound the query complexity of evaluation of a relation.  
The lower bound is given by the following optimisation problem:

\begin{subequations}
\label{eqn:relationDual}
\begin{alignat}{3}
&\mbox{\rm minimise} &\quad& \max \sfigB{ \max\nolimits_{x\in \cD} \norm|u_{x}|^2, \max\nolimits_{y\in\cD} \norm|v_{y}|^2 } \\
& \mbox{\rm subject to}&&  
1 - \sum\nolimits_{a\in [m]} \ip<\sigma_{x,a},\sigma_{y,a}> = \ip< u_x, \sA[\Delta_{xy}\otimes I_{\cW}] v_y> && \text{\rm for all $x,y\in\cD$;} \label{eqn:relationDualCondition} \\
&&& \sum\nolimits_{a\notin r(x)} \|\sigma_{x,a}\|^2 \le \eps && \text{\rm for all $x\in\cD$;}\\
&&& \text{$\cH,\cW$ are vector spaces,\qquad $u_x, v_y \in \cX\otimes\cW$,\qquad $\sigma_{x,a}\in\cH$.
\!\!\!\!\!\!\!\!\!\!\!\!\!\!\!\!
}
\end{alignat}
\end{subequations}
Indeed, this is the adversary bound~\rf(eqn:advExplicit) with the optimisation over $\sigma_x = \bigoplus_{a\in[m]} \sigma_{x,a}$.  The condition~\rf(eqn:relationDualCondition) with $x=y$ implies $\|\sigma_x\|^2 = 1$.
Taking the dual, we obtain the following formulation:
\begin{subequations}
\label{eqn:relationApproximate}
\begin{alignat}{3}
&\mbox{\rm maximise} &\quad& \lambda_{\max} \sA[\Gamma - \eps \mathrm N] \\
& \mbox{\rm subject to}&&  
\norm| \Gamma\circ\Delta | \le 1\;; \label{eqn:relationApproximateCondition}\\
&&& \Gamma \preceq \mathrm N\circ \mathrm E_a &\quad& \text{\rm for all $a\in[m]$.}
\end{alignat}
\end{subequations}
Here $\Gamma$ is a Hermitian $\cD\times \cD$ matrix, and $\mathrm N$ is a diagonal $\cD\times \cD$ matrix with non-negative real entries.  Next, $\lambda_{\max}$ stands for the maximal eigenvalue, 
thus, $\|\Gamma\|=\max\{\lambda_{\max}(\Gamma),\lambda_{\max}(-\Gamma)\}$.
The diagonal $\cD\times \cD$ matrix $\mathrm E_a$ is defined by $\mathrm E_a\elem[x,x]=1_{a\notin r(x)}$.
The Hadamard product in~\rf(eqn:relationApproximateCondition) is defined as in \rf(prp:gamma2dual).

When $\eps=0$, the optimisation problem can be slightly simplified (where $\Gamma\elem[r^{-1}(a),r^{-1}(a)]$ stands for the corresponding submatrix of $\Gamma$):
\begin{subequations}
\label{eqn:relationExact}
\begin{alignat}{3}
&\mbox{\rm maximise} &\quad& \lambda_{\max} \sA[\Gamma] \\
& \mbox{\rm subject to}&&  
\norm| \Gamma\circ\Delta | \le 1 \;;\label{eqn:relationExactCondition}\\
&&& \Gamma \elem[r^{-1}(a), r^{-1}(a) ] \preceq 0 &\quad& \text{\rm for all $a\in[m]$.} \label{eqn:relationExactSub}
\end{alignat}
\end{subequations}
(Compare condition~\rf(eqn:relationExactSub) with condition~\rf(eqn:functionExactGamma) from the functional case.)
The dual formulations in~\rf(eqn:relationApproximate) and~\rf(eqn:relationExact) are derived in Appendices~\ref{app:relation} and~\ref{app:relationExact}.  Thus, we obtain:

\begin{thm}
A lower bound on evaluating a relation $r$ with input oracles $\{O_x\}$ and error $\eps$ is given by~\rf(eqn:relationApproximate).  A lower bound on exact evaluation is given by~\rf(eqn:relationExact).
\end{thm}

By \rf(thm:conversionUpper), the approximate bound is essentially tight if one can tolerate a slightly larger error in the upper bound.

We can also get a lower bound for the \emph{average-case} error.
 In this case, a probability distribution $p_x$ is given on $\cD$, and the algorithm must have error at most $\eps$ on average, when the input $x$ is sampled from $p$.  
In notation of \rf(defn:relation), the requirement is $\sum_{x\in\cD} p_x \| (\Pi_{r(x)}\otimes I_{\cH})\sigma_x\|^2 \ge 1-\eps$.
The following result is proven in \rf(app:average).

\begin{thm}
\label{thm:averageError}
The quantum query complexity of evaluating a relation $r$ with input oracles $\{O_x\}$, probability distribution $p_x$ on the set of inputs $\cD$, and average error $\eps$, is lower bounded by the following optimisation problem:
\begin{subequations}
\label{eqn:averageError}
\begin{alignat}{3}
&\mbox{\rm maximise} &\quad& u^*\Gamma u - \eps\eta \\
& \mbox{\rm subject to}&&  
\norm| \Gamma\circ\Delta | \le 1\;;\label{eqn:averageErrorCondition} \\
&&& \Gamma\circ(uu^*) \preceq \eta P \circ \mathrm E_a &\quad& \text{\rm for all $a\in[m]$;}
\end{alignat}
\end{subequations}
where the optimisation is over $\cD\times\cD$ Hermitian matrix $\Gamma$, unit vector $u\in\bC^{\cD}$, and real $\eta\ge 0$.  The matrix $P=\mathrm{diag}(p_x)$, and $\Delta$ and $\mathrm E_a$ are as in~\rf(eqn:relationApproximate).
\end{thm}

Now assume we have the standard input oracle as in \rf(defn:standardOracle).
Without loss of generality, without loss of generality $\cD = [q]^n$, as we may always say that $x\in[q]^n\setminus\cD$ is in relation with all $a\in[m]$.
Next, by \rf(prp:preparationOrhtogonal), we may replace $\Delta_{xy}$ in~\rf(eqn:relationDualCondition) by $\bigoplus_{j\in[n]} 1_{x_j\ne y_j}$.  Using the same dual, and simplifying slightly, we obtain the following version of the bound.

\begin{cor}
Let $r\colon [q]^n\to[m]$ be a relation.  
The quantum query complexity of evaluating $r$ with the standard input oracle is given by~\rf(eqn:relationApproximate) and~\rf(eqn:relationExact) for the approximate and exact versions of the problem, respectively, where the conditions~\rf(eqn:relationApproximateCondition) and~\rf(eqn:relationExactCondition) are replaced by
\[
\| \Gamma\circ\Delta_j\|\le 1\qquad\text{\rm for all $j\in[n]$},
\]
and the $[q]^n\times [q]^n$ matrix $\Delta_j$ is given by $\Delta_j\elem[x,y] = 1_{x_j\ne y_j}$.
The complexity in the average-error case is given by~\rf(eqn:averageError) with~\rf(eqn:averageErrorCondition) replaced by the same condition.
\end{cor}

\subsection{Purifiers}
\label{sec:purifiers}

As we mentioned in the introduction, the usual adversary bound is tight for bounded-error function evaluation, but it is only semi-tight for general state conversion problems.  In this section, we give an explanation of this phenomenon and show that this is still true for general input oracles.
Our method can be extended to other problems.
For instance, we show that the adversary bound is tight for approximate evaluation of relations, provided that we can efficiently test the output of the algorithm.

The functional case is based on the following state generation problem.
Let $\cY$ and $\cH$ be vector spaces.
The input vector $\psi_x\in \cY\otimes\cH$, when $\cY$ is measured, gives one of the basis vector with high probability.  The target vector is the corresponding basis vector.
We show that if the problem has a bounded gap, its adversary bound is constant.  
More precisely, we consider the following two cases:
\itemstart
\item Let $n=2$, and the standard basis of $\cY$ be $\{e_0,e_1\}$.
Assume there exist constants $c,\delta$ with $0<c-\delta<c+\delta<1$, such that, for each input vector $\psi_x$, either $\|\Pi_1 \psi_x\|^2 \le c-\delta$, or $\|\Pi_1 \psi_x\|^2 \ge c+\delta$, where $\Pi_i = (e_ie_i^*)\otimes I_{\cH}$.  In the first case we define $\sigma_x = e_0$, and in the second case $\sigma_x = e_1$.
\item Let $n$ be arbitrary, $\cY$ have the standard basis $\{e_1,\dots,e_n\}$, and $\delta>0$ be a constant.
For each $x\in\cD$, there exists $i\in[n]$ such that $\|\Pi_i \psi_x\|^2 \ge 1/2 + \delta$.  The corresponding target state is $\sigma_x = e_i$.
\itemend

\begin{thm}
\label{thm:purifierFunction}
For both of the above state generation problems, we have 
\[
\gamma_2 \sA[1-\ip<\sigma_x, \sigma_y> \mid 1-\ip<\psi_x,\psi_y> ]_{x,y\in\cD} = O(1).
\]
\end{thm}
Note that in this case $1-\ip<\sigma_x,\sigma_y> = 1_{\sigma_x\ne\sigma_y}$.
We give the proof at the end of this section.
Depending on the point of view, this result can be interpreted in two ways.

\begin{cor}
Let $O_x$ be the standard input oracle encoding a string $x\in [q]^n$ as in \rf(defn:standardOracle).
Assume $\circuit A$ is a quantum algorithm implementing a unitary $V_x$ in $T$ queries to $O_x$ for all $x\in\cD\subseteq[q]^n$.
Then, the unitary $V_x$ can be approximated to arbitrary constant precision in $O(T)$ queries to a noisy oracle $\tO_x$.
\end{cor}

A usual implementation of $V_x$ with the noisy oracle reduces the error of the noisy oracle to $\ll 1/T$ by $O(\log T)$ repetitions, giving an algorithm with complexity $O(T\log T)$.  The previous corollary removes this logarithmic factor.

\pfstart
By \rf(thm:unitaryLower) and \rf(prp:preparationOrhtogonal), we have
\[
\gamma_2\sB[V_x - V_y \midB \bigoplus\nolimits_{j\in[n]} 1_{x_j \ne y_j}]_{x,y\in\cD} \le 2T.
\]
On the other hand, the states $\psi_{x,j}$ generated by the noisy oracle $\tO_x$ satisfy the conditions of \rf(thm:purifierFunction), so using this theorem and \rfitem(direct@prp:gamma2prop),
\[
\gamma_2\sB[\bigoplus\nolimits_{j\in[n]} 1_{x_j \ne y_j} \midB \bigoplus\nolimits_{j\in[n]} (1-\ip<\psi_{x,j}, \psi_{y,j}>)]_{x,y\in\cD} = O(1).
\]
The statement now follows from \rfitem(composition@prp:gamma2prop), \rf(prp:preparationEasy) and \rf(thm:unitaryUpper).
\pfend

Another implication of \rf(thm:purifierFunction) is our promised result about the tightness of the adversary bound for bounded-error function evaluation.

\begin{thm}
\label{thm:advFunction}
Let $f\colon \cD\to [m]$ be a function, and $\{O_x\}_{x\in\cD}$ be a collection of input oracles.
The adversary bound
\begin{equation}
\label{eqn:advFunction}
\gamma_2 \sA[ 1_{f(x)\ne g(x)} \mid O_x - O_y]_{x,y\in\cD}
\end{equation}
characterises the bounded-error quantum query complexity of evaluating $f$ with the general oracle $\{O_x\}_{x\in\cD}$ up to a constant factor.
\end{thm}

\pfstart
The adversary bound gives an upper bound by \rf(thm:conversionUpper).
For the lower bound, let $\circuit A$ be an optimal algorithm that evaluates $f$ with bounded error in $T$ queries.
Let $\psi_x$ be the final state of $\circuit A$ on the input $x\in\cD$.
By \rf(thm:conversionLower),
\[
\gamma_2\sA[ 1-\ip<\psi_x,\psi_y> \mid O_x - O_y ]_{x,y\in\cD} \le T.
\]
The vectors $\psi_x$ satisfy the conditions of \rf(thm:purifierFunction), hence,
\[
\gamma_2\sA[ 1_{f(x)\ne f(y)} \mid 1-\ip<\psi_x,\psi_y> ]_{x,y\in\cD} = O(1) .
\]
Using \rfitem(composition@prp:gamma2prop), we get the statement of the theorem.
\pfend

Applying \rf(prp:gamma2dual) to~\rf(eqn:advFunction) and simplifying (or using~\rf(eqn:relationExact)), we get the following formulation of the adversary bound for function evaluation:
\begin{subequations}
\label{eqn:functionExact}
\begin{alignat}{3}
&\mbox{\rm maximise} &\quad& \|\Gamma\| \\
& \mbox{\rm subject to}&&  
\norm| \Gamma\circ\Delta | \le 1\;; \\
&&& \Gamma \elem[f^{-1}(a), f^{-1}(a) ] = 0 &\quad& \text{\rm for all $a\in[m]$;} \label{eqn:functionExactGamma}
\end{alignat}
\end{subequations}
with notation as in~\rf(eqn:relationApproximate).
\medskip

In general, we cannot obtain an analogue of \rf(thm:purifierFunction) for approximate evaluation of relations.
The reason is ambiguity.  Assume there is an element $x\in\cD$ in relation with all $[m]\setminus\{1\}$, and an element $y\in\cD$ in relation with all $[m]\setminus\{2\}$.
Given a uniform superposition over all elements of $[m]$ as the input state, should we amplify for the first or for the second case?

But assume we can effectively \emph{test} a solution for a relation $r$.  That is, there exists a bounded-error quantum algorithm that, given oracle access to $O_x$, and an element $a\in [m]$, accepts iff $a\in r(x)$.  
This is a common scenario, especially in so-called {\em search problems}.
For instance, consider the relation $r$ between $[q]^n$, where $n\gg q$, and the set of 2-subsets of $[n]$.  A string $x\in [q]^n$ is in relation with $\{a,b\}\subset [n]$ iff $x_a = x_b$.  If $q$ is large, it is not easy to find a pair $\{a,b\}$ in relation with $x$, but, having a pair, it is trivial to test it.
Having this situation in mind, we state the following result.

\begin{thm}
\label{thm:purifierGeneral}
Assume $\cZ$ is a vector space, and $\delta>0$.  Let $\{\psi_x,\Pi_x\}_{x\in\cD}$ be a collection of unit vectors and projectors in $\cZ$ such that $\|\Pi_x\psi_x\|^2 \ge \delta$ for all $x$.
For each $x$, let $R_x$ be the reflection $I_\cZ-2\Pi_x$.
Then, there exist a vector space $\cH$, and a collection $\{\sigma_x\}_{x\in\cD}$ of unit vectors in $\cZ\otimes \cH$ such that $(\Pi_x\otimes I_\cH) \sigma_x = \sigma_x$ for all $x$, and
\[
\gamma_2\sB [ 1-\ip<\sigma_x , \sigma_y> \midA (1-\ip<\psi_x,\psi_y>)\oplus(R_x - R_y) ]_{x,y\in\cD} \le \frac2\delta\;.
\]
\end{thm}

We will give the proof later.  Actually, it is easy to improve the bound slightly.
\begin{cor}
In the settings of \rf(thm:purifierGeneral), let $O_x$ be an oracle generating the state $\psi_x$.
Then,
\begin{equation}
\label{eqn:purifierCor}
\gamma_2\sB [ 1-\ip<\sigma_x , \sigma_y> \midA (O_x-O_y)\oplus(R_x - R_y) ]_{x,y\in\cD} = O\s[\frac1{\sqrt{\delta}}]\;.
\end{equation}
\end{cor}

\pfstart
By amplitude amplification~\cite{brassard:amplification}, in $O(1/\sqrt{\delta})$ queries to $O_x$, we can generate a state $\psi'_x \in \cZ\otimes \cH$ such that $\|(\Pi_x\otimes I_\cH)\psi'_x\|^2 = \Omega(1)$.  
By \rf(thm:conversionLower), and Points~(\ref{reflexivity@prp:gamma2prop}) and~(\ref{direct@prp:gamma2prop}) of \rf(prp:gamma2prop), we have
\[
\gamma_2\sB [ (1-\ip<\psi_x' , \psi_y'>)\oplus(R_x - R_y) \midA (O_x-O_y)\oplus(R_x - R_y) ]_{x,y\in\cD} = O\s[\frac1{\sqrt{\delta}}]\;.
\] 
Now compose it, using \rfitem(composition@prp:gamma2prop), with the bound from~\rf(thm:purifierGeneral) applied to the vectors $\psi'_x$.
\pfend

The $O_x-O_y$ term in~\rf(eqn:purifierCor) cannot be replaced with $1-\ip<\psi_x,\psi_y>$ due to the reasons explained in \rf(sec:counterexample).

Our main consequence of \rf(thm:purifierGeneral) is that, in this case, it is possible to replace bound~\rf(eqn:relationApproximate) with an easier bound~\rf(eqn:relationExact), when analysing approximate evaluation of relations.

\begin{thm}
Let $r\colon \cD\to [m]$ be a relation, and $\{O_x\}_{x\in\cD}$ be a collection of input oracles. Let $P$ be the value of the adversary bound~\rf(eqn:relationExact),
which also equals~\rf(eqn:relationDual) with $\eps=0$.
Assume that there exists a bounded-error quantum algorithm that, given oracle access to $O_x$ and $a\in [m]$, tests whether $a\in r(x)$ in $o(P)$ queries.
Then, $\Omega(P)$ is a lower bound on the quantum query complexity of evaluating $r$ with general input oracles $\{O_x\}$ and error $1-\Omega(1)$.
\end{thm}

\pfstart
The proof is similar to the one of \rf(thm:advFunction), but this one is a slightly more complicated jigsaw puzzle.

Let $\circuit A$ be an optimal algorithm that evaluates $r$ with error $1-\delta$ in $T$ queries, where $\delta = \Omega(1)$.  Our aim is to show that $T = \Omega(P)$.
Let $\cY\otimes \cH$ be the space of the algorithm $\circuit A$, where $\cY = \bC^m$.
Let $\psi_x\in \cY\otimes \cH$ be the final state of $\circuit A$ on the input $x\in\cD$.
By \rf(thm:conversionLower),
\begin{equation}
\label{eqn:advRelation1}
\gamma_2\sA[ 1-\ip<\psi_x,\psi_y> \mid O_x - O_y ]_{x,y\in\cD} \le T.
\end{equation}
Now we apply \rf(thm:purifierGeneral) with $\Pi_x = \Pi_{r(x)}\otimes \cI_{\cH}$, where $\Pi_{r(x)}$ is as in \rf(defn:relation).  This gives
\begin{equation}
\label{eqn:advRelation2}
\gamma_2\sB [ 1-\ip<\sigma_x , \sigma_y> \midA (1-\ip<\psi_x,\psi_y>)\oplus(R_x - R_y) ]_{x,y\in\cD} = O(1)
\end{equation}
for some vectors $\sigma_x$ satisfying the constraints of \rf(thm:purifierGeneral).

For $a\in[m]$, let $f_a\colon \cD\to \{0,1\}$ be defined by $f_a(x) = 1_{a\in r(x)}$.
For the reflections $R_x$ in~\rf(eqn:advRelation2), we have
\[
R_x - R_y = \sC[\bigoplus\nolimits_{a\in[m]} 2 (-1)^{f_a(x)}\cdot 1_{f_a(x)\ne f_a(y)} ]\otimes I_{\cH}.
\]
We know that each $f_a$ can be evaluated in some number $S = o(P)$ queries to $O_x$, hence,
by \rf(thm:advFunction) and \rfitem(direct@prp:gamma2prop),
\[
\gamma_2\sC[\bigoplus\nolimits_{a\in[m]} 1_{f_a(x)\ne f_a(y)} \midB O_x - O_y]_{x,y,\in\cD} = O(S).
\]
Hence, using Points~(\ref{tensor@prp:gamma2prop}) and~(\ref{linear@prp:gamma2prop}) of \rf(prp:gamma2prop), we have
\(
\gamma_2\sA[R_x - R_y \mid O_x - O_y]_{x,y,\in\cD} = O(S).
\)
Combining this with~\rf(eqn:advRelation1) using \rfitem(direct@prp:gamma2prop), we arrive at
\[
\gamma_2\sB[(1-\ip<\psi_x,\psi_y> )\oplus (R_x - R_y) \mid O_x - O_y]_{x,y,\in\cD} = O(S+T).
\]
Finally, combining this with~\rf(eqn:advRelation2) using \rfitem(composition@prp:gamma2prop), we get 
\[
\gamma_2\sA[ 1-\ip<\sigma_x,\sigma_y> \mid O_x - O_y]_{x,y,\in\cD} = O(S+T).
\]
The left-hand side gives a feasible solution to the optimisation problem in~\rf(eqn:relationDual) with $\eps=0$.
Since $P$ is defined as the optimal solution to that problem, $P=O(S+T)$, and as $S = o(P)$, we get $T = \Omega(P)$.
\pfend

We now prove both our technical results.

\pfstart[Proof of \rf(thm:purifierGeneral)]
Our construction is inspired by the following simple algorithm:
Generate the state $\psi_x$, then, conditioned on being orthogonal to $\Pi_x$, generate another copy of $\psi_x$, and repeat.  After infinitely many repetitions, we obtain a state $\sigma_x$ satisfying the constraints.  But as the ``bad'' part of the state becomes exponentially small through the course of the algorithm, we manage to improve the complexity from $+\infty$ to $2/\delta$.

It is most convenient to describe the construction using infinite-dimensional $\sigma_x$.  Namely, we take
\[
\sigma_x = (\Pi_x\psi_x)\otimes \sC[\bigoplus_{k=0}^\infty\; (\Pi_x^\perp \psi_x)^{\otimes k}],
\]
where $\Pi_x^\perp = I_\cX - \Pi_x$.  
This corresponds to the final state of the above infinite algorithm up to rearrangement of the registers.
We have
\begin{equation}
\label{eqn:sigmaxsigmay}
\ip<\sigma_x,\sigma_y> = \ip<\Pi_x\psi_x,\Pi_y\psi_y>\sum_{k=0}^\infty \ipA<\Pi_x^\perp\psi_x,\Pi_y^\perp\psi_y>^k = \frac{\ip<\Pi_x\psi_x,\Pi_y\psi_y>}{1 - \ip<\Pi_x^\perp\psi_x,\Pi_y^\perp\psi_y>}\;.
\end{equation}
In particular, $\|\sigma_x\| = 1$.  In order to proceed, we need the following technical result.

\begin{lem}
\label{lem:1-Xij}
Assume $X$ is a matrix with $\gamma_2(X)<1$.
Let $Y$ be the matrix of the same size defined by $Y\elem[i,j] = 1/(1-X\elem[i,j])$.
Then, $\gamma_2(Y)\le 1/(1-\gamma_2(X))$.
\end{lem}

\pfstart
By \rfitem(reflexivity@prp:gamma2prop), $|X\elem[i,j]|< 1$ for all $i$ and $j$, hence, the matrix $Y$ is well-defined.
Then, using the triangle inequality and \rfitem(hadamard@prp:gamma2prop),
\[
\gamma_2(Y) = \gamma_2\sC[\sum_{k=0}^\infty X{\elem[i,j]}^k]_{i,j} \le \sum_{k=0}^\infty \gamma_2(X)^k = \frac1{1-\gamma_2(X)}.\qedhere
\]
\pfend

As $\gamma_2\sA[ \ipA<\Pi_x^\perp\psi_x,\Pi_y^\perp\psi_y> ]_{x,y\in\cD} \le 1-\delta$, we get from \rf(lem:1-Xij) that
\begin{equation}
\label{eqn:purifier1}
\gamma_2\s[ \frac1{1- \ipA<\Pi_x^\perp\psi_x,\Pi_y^\perp\psi_y>} ]_{x,y\in\cD} \le \frac1\delta.
\end{equation}
Next, it is easy to check that
\[
1 - \ipA<\Pi_x\psi_x, \Pi_y\psi_y> - \ipA<\Pi_x^\perp\psi_x, \Pi_y^\perp \psi_y>
= \skA[1-\ip<\psi_x, \psi_y>] + (\Pi_x\psi_x - \Pi_x^\perp \psi_x)^* \skA[\Pi_x - \Pi_y] \psi_y,
\]
hence, using that $\Pi_x - \Pi_y = -\frac12 (R_x - R_y)$,
\begin{equation}
\label{eqn:purifier2}
\gamma_2\s[ 1 - \ipA<\Pi_x\psi_x, \Pi_y\psi_y> - \ipA<\Pi_x^\perp\psi_x, \Pi_y^\perp \psi_y>
\midA (1-\ip<\psi_x,\psi_y>)\oplus(R_x - R_y) ]_{x,y\in\cD} \le 2.
\end{equation}
Hence, combining~\rf(eqn:purifier1) and~\rf(eqn:purifier2) via \rfitem(hadamard@prp:gamma2prop), and using~\rf(eqn:sigmaxsigmay), we get
\[
\gamma_2\sB [ 1-\ip<\sigma_x , \sigma_y> \midA (1-\ip<\psi_x,\psi_y>)\oplus(R_x - R_y) ]_{x,y\in\cD} \le \frac2\delta.
\]

It remains to explain how to make $\sigma_x$ finite-dimensional.
Recall our assumption on $\cD$ being finite.  
Using a standard dimensional argument, there exist vectors $\tau_x\in \bC^{\cD}$ such that
\[
\ip<\tau_x, \tau_y> =  
\ipC<\bigoplus_{k=0}^\infty\; (\Pi_x^\perp \psi_x)^{\otimes k},\; \bigoplus_{k=0}^\infty\; (\Pi_y^\perp \psi_y)^{\otimes k}>
\]
for all $x,y\in\cD$.  Thus, we can take $\sigma_x = (\Pi_x\psi_x)\otimes\tau_x$.
\pfend

\pfstart[Proof of \rf(thm:purifierFunction)]
Using an analogue of~\rf(eqn:1dgamma2), we see that it suffices to show that
\[
\gamma_2\s[\frac{1_{\sigma_x\ne\sigma_y}}{1-\ip<\psi_x,\psi_y>}]_{x,y\in\cD} = O(1),
\]
where $0/0=0$.
It is not hard to show that $\gamma_2(1_{\sigma_x\ne\sigma_y})_{x,y\in\cD} \le 2$.
Thus, by \rfitem(hadamard@prp:gamma2prop), it suffices to prove that $\gamma_2(Y) = O(1)$ for some matrix $Y$ such that $Y\elem[x,y] = 1/(1-\ip<\psi_x,\psi_y>)$ for all $x,y$ satisfying $\sigma_x\ne\sigma_y$.
By \rf(lem:1-Xij), it suffices to find a $\cD\times\cD$ matrix $X$ such that $X\elem[x,y]=\ip<\psi_x,\psi_y>$ whenever $\sigma_x\ne\sigma_y$, and $\gamma_2(X) = 1-\Omega(1)$.
\medskip

In the first case of the state conversion problem, we define $X\elem[x,y] = \ip<\phi_x, \phi_y>$, where $\phi_x\in\cX\otimes\cH$ is given by
\begin{equation}
\label{eqn:phix}
\phi_x = \begin{cases}
\sqrt{\alpha}\; \Pi_0 \psi_x + \sqrt{\beta}\; \Pi_1 \psi_x, & \text{if $\|\Pi_1 \psi_x\|^2\le c-\delta$;}\\
\frac1{\sqrt{\alpha}}\; \Pi_0 \psi_x + \frac1{\sqrt{\beta}}\; \Pi_1 \psi_x, & \text{if $\|\Pi_1 \psi_x\|^2\ge c+\delta$;}\\
\end{cases}
\end{equation}
for
\[
\alpha = \sqrt{\frac{1-c-\delta}{1-c+\delta}}\qquad\text{and}\qquad \beta = \sqrt{\frac{c+\delta}{c-\delta}}.
\]

It is clear that $\ip<\phi_x,\phi_y> = \ip<\psi_x,\psi_y>$ whenever $\sigma_x\ne\sigma_y$,
i.e., the definitions of $\phi_x$ and $\phi_y$ are given by different cases in \rf(eqn:phix).
Now assume $\|\Pi_1 \psi_x\|^2 = c - \delta - s$ for some $s\ge 0$.  Then,
\[
\|\phi_x\|^2 = (1-c +\delta+s) \alpha + (c - \delta - s)\beta
\le	(1-c+\delta)\alpha + (c-\delta)\beta = \sqrt{(1-c)^2-\delta^2} + \sqrt{c^2 - \delta^2}<1.
\]
Similarly, if $\|\Pi_1 \psi_x\|^2 = c + \delta +s$ for some $s\ge 0$, then
\[
\|\phi_x\|^2 = \frac{1-c -\delta-s}{\alpha} + \frac{c + \delta + s}\beta
\le \frac{1-c -\delta}{\alpha} + \frac{c + \delta}\beta
= \sqrt{(1-c)^2-\delta^2} + \sqrt{c^2 - \delta^2}<1.
\]
Thus, $\gamma_2(X)$ is bounded away from 1, and this finishes the proof for the first case.
\medskip

The second case is similar.  Now $\cX$ has the standard basis $\{e_1,\dots,e_n\}$, and let $\cY$ be a space with the standard basis $\{e_0,e_1,e_2\}$.
We define $X\elem[x,y] = \ip<u_x,v_y>$, where $u_x,v_x \in\cY\otimes\cX\otimes\cH$ are defined as follows.
For a fixed $x$, let $i\in[n]$ be such that $\|\Pi_i\psi_x\|^2\ge 1/2+\delta$.
Then,
\[
u_x = \frac{\ket Y|0>+\sqrt{2\delta}\;\ket Y|1>}{1+2\delta}\otimes \ket {\cX\otimes \cH}|\Pi_i \psi_x>
+ \sum_{j\in[n]\setminus \{i\}} \sB[\ket Y|0>+\sqrt{2\delta}\;\ket Y|2>]\otimes \ket \cX\otimes\cH|\Pi_j \psi_x>,
\]
and
\[
v_x = \frac{\ket Y|0>+\sqrt{2\delta}\;\ket Y|2>}{1+2\delta}\otimes \ket {\cX\otimes \cH}|\Pi_i \psi_x>
+ \sum_{j\in[n]\setminus \{i\}} \sB[\ket Y|0>+\sqrt{2\delta}\;\ket Y|1>]\otimes \ket \cX\otimes\cH|\Pi_j \psi_x>,
\]
It is not hard to check that $\ip<u_x,v_y>=\ip<\psi_x,\psi_y>$ whenever $\sigma_x\ne\sigma_y$.
Also, both $\|u_x\|^2$ and $\|v_x\|^2$ are bounded by
\[
\frac1{1+2\delta} \|\Pi_i\psi_x\|^2 +  (1+2\delta) \|\psi_x - \Pi_i\psi_x\|^2 \le
\frac{1/2+\delta}{1+2\delta} + (1+2\delta)(1/2-\delta) = 1-2\delta^2.
\]
Thus, $\gamma_2(X)$ is bounded away from 1, and the second case is proven.
\pfend

\section{Future Directions}
In this section, we mention some possible future research directions.

\begin{itemize}
\item
Besides the \emph{additive} quantum adversary, which we studied in this paper, \v Spalek~\cite{spalek:multiplicative} developed \emph{multiplicative} quantum adversary.
Ambainis \etal~\cite{ambainis:symmetryAssisted} generalised it to the state generation problem.
However, it would be interesting to understand whether the multiplicative adversary can be obtained in the style of~\cite{lee:stateConversion}, i.e., as a relaxation of a quantum algorithm.  Can any of the theory developed in this paper be applied to it?  Are there other possible relaxations worth studying?

\item
In \rf(sec:counterexample), we showed that the adversary bound for unitary implementation is more precise than the adversary bound for state conversion.
How big is the separation?  Can more convenient forms of \rf(prp:preparationHard) be obtained?

\item
What are the conditions when effective purifiers exist?  In other words, when does the adversary bound provide a lower bound not only for the exact, but also for the approximate version of the problem?

\item
Finally, it would be interesting to obtain some applications for the theory presented in this paper, particularly, for the bounds developed in \rf(sec:relationBasic).
\end{itemize}

One theoretical problem, for which these tools may be appropriate, is development of the quantum theory for composition of relations.  
We briefly define the problem, for simplicity focusing on a special case of evaluating a function $f\colon [q]^n\to [m]$, when superpositions are allowed as inputs.
The algorithm is given an oracle $O_x$ that generates states $\psi_{x,j}\in \cY\otimes \cH$, where $\cY = \bC^q$.  For $X\subseteq[q]$, let $\Pi_X = \sum_{a\in X} e_ae_a^*$.

\mycommand{tf}{\tilde f}
\mycommand{tor}{\widetilde{\mathrm{OR}}}
The function is given as a sequence of rules 
$
\tf(X_1,\dots,X_n) = a
$ with $X_j\subseteq[q]$ and $a\in[m]$.
Each of these rules means that the algorithm must output $a$ whenever $(\Pi_{X_j}\otimes I_\cH) \psi_{x,j} = \psi_{x,j}$ for all $j\in[n]$.
For example, for the OR function, we may consider the rules of the form
\begin{equation}
\label{eqn:ORrules}
\tor\sA[\{0,1\},\dots,\{0,1\}, \{1\}, \{0,1\},\dots,\{0,1\}] = 1,\qquad\text{and}\qquad
\tor\sA[\{0\},\dots,\{0\}] = 0.
\end{equation}

For randomised algorithms, this problem is equivalent to the evaluation of the underlying usual function, since, when an input variable is accessed, it ``collapses'' to one of the elements of $[q]$.  Quantumly, the situation is very different.  If the algorithm works with the oracle generating the state $\ket |0>$ and the oracle generating $\ket |1>$, this does not mean it works with an oracle that prepares some superposition of the two states.

As a lower bound for this problem, it is possible to take the adversary bound for the underlying usual function, i.e., when the input oracle can only generate basis states.
From~\rf(eqn:advOrig), it is easy to obtain the following upper bound:
\begin{subequations}
\label{eqn:overlap}
\begin{alignat}{3}
&\mbox{\rm minimise} &\quad& \max_{X} \max \left\{ \sum\nolimits_{j\in [n]}\|u_{X,j}\|^2\right., \!\!\!\! && \;\;\left.\sum\nolimits_{j\in [n]}\|v_{X,j}\|^2 \right\}  \\
& \mbox{\rm subject to}&& \sum\nolimits_{j\in[n]} \ip<u_{X,j}, v_{Y,j}> = 1 && \text{\rm whenever $f(X)\ne f(Y)$;}  \\
&&& \ip<u_{X,j}, v_{Y,j}> = 0 && \text{\rm whenever $X_j\cap Y_j \ne \emptyset$;}
\end{alignat}
\end{subequations}
where $X$ and $Y$ range over all rules of the function $f$.
For some functions, the bound in~\rf(eqn:overlap) can be obtained for free from the usual adversary bound.  For instance, from a solution to the adversary bound for the OR function, we get an algorithm for $\tor$ given by~\rf(eqn:ORrules).  It would be interesting to get a precise characterisation of the complexity of this problem.

\subsection*{Acknowledgements}
I am grateful to Robin Kothari for various discussions and many useful suggestions, as well as for sharing his manuscript~\cite{kothari:SDPCharacterization}.  I thank Troy Lee for useful comments.

The author is supported by FP7 FET Proactive project QALGO.
Parts of this work was done while at CSAIL, Massachusetts Institute of Technology, USA, supported by Scott Aaronson's Alan T. Waterman Award from the National Science Foundation, and at CQT, Singapore.  The author thanks Miklos Santha for hospitality.

\bibliographystyle{\relativepath habbrvM}
{
\bibliography{../../bib}
}


\appendix

\section{Duality}
In this appendix, we derive all the dual formulations stated in the paper.
Our treatment of convex duality is based on~\cite[Chapter 5]{boyd:convex}.  We construct the dual by explicitly writing down the Lagrangian and transforming it.  Thus, \emph{weak duality} (the maximisation problem bounds the minimisation problem from below) is apparent.  
To prove \emph{strong duality} (their optimal values are equal), we rely on Slater's condition.  The latter says that strong duality holds if one of the optimisation problems is convex and \emph{strictly feasible}, i.e. there exists a feasible solution making all the inequalities in the problem strict.

We assume the reader is familiar with basic properties of semidefinite matrices.  We use notation $\ip<A,B> = \tr A^*B$ for the inner product of matrices.  The main property we use is that if $A$ is Hermitian, then $\ip<X,A>\ge 0$ for all $X\succeq 0$ if and only if $A\succeq 0$.

\subsection{Proof of \rf(prp:gamma2dual)}
\label{app:Gamma2Scalar}

Let us restate the problem for the case when all $A_{xy}$ are one-dimensional.

\begin{alignat*}{2}
&\mbox{\rm minimise} &\quad& \max \sfigB{ \max\nolimits_{x\in \cD_1} \norm|u_{x}|^2, \max\nolimits_{y\in\cD_2} \norm|v_{y}|^2 }  \\
& \mbox{\rm subject to}&&  
a_{xy} = \ip< u_x, \sA[\Delta_{xy}\otimes I_{\cW}] v_y> \qquad \text{\rm for all $x\in\cD_1$ and $y\in\cD_2$;}  \\
&&& \text{$\cW$ is a vector space,\qquad $u_x \in \cX_1\otimes\cW$,\qquad $v_y\in \cX_2\otimes\cW$.}
\end{alignat*}

If there exist $x$ and $y$ such that $\Delta_{xy}=0$ but $a_{xy}\ne 0$, then both formulations of the problem have optimal value $+\infty$, so we may assume that $a_{xy}=0$ whenever $\Delta_{xy}=0$.

Assume that $\cD_1$ and $\cD_2$ are disjoint, and let $\ctD = \cD_1\cup\cD_2$.
We may decompose $u_x = \bigoplus_i u_{x,i}$ and $v_y = \bigoplus_j v_{y,j}$, where $u_{x,i}, v_{y,j}\in\cW$, and $i$ and $j$ run through the elements of the bases of $\cX_1$ and $\cX_2$, respectively.
Let $X$ be the Gram matrix of the vectors $\{u_{x,i}\}$ and $\{v_{y,j}\}$.  
We treat $X$ as a $\ctD\times\ctD$ block-matrix with blocks $X_{xy}$.
That is, if $x,y\in\cD_1$, then $X_{xy}\colon \cX_1\to\cX_1$ is given by $X_{xy}\elem[i,j] = \ip<u_{x,i}, u_{x,j}>$.  The remaining three cases $x\in\cD_1$, $y\in\cD_2$; $x\in\cD_2$, $y\in\cD_1$; and $x,y\in\cD_2$ are similar.
With this change, the optimisation problem reads as follows:
\begin{subequations}
\label{eqn:dualGamma2Scalar}
\begin{alignat}{3}
&\mbox{\rm minimise} &\quad& t \\
& \mbox{\rm subject to}&&  \tr X_{zz} \le t &\quad& \text{for all $z\in\ctD$\;;} \\
&&& a_{xy} = \ip<X_{xy}, \Delta_{xy}> && \text{for all $x\in\cD_1$ and $y\in\cD_2$;} \label{eqn:dualGamma2ScalarCondition}\\
&&& t\in\bR,\quad X\succeq 0.
\end{alignat}
\end{subequations}

We have the following Lagrangian
\[
L(t, X, \mu, \lambda) = t + \sum_{z\in\ctD} \mu_z (\tr X_{zz} - t) + \sum_{x\in\cD_1, y\in\cD_2} 2\mathrm{Re} \sB[{ \lambda_{xy} \sA[ a_{xy} - \ip<X_{xy}, \Delta_{xy}>]}],
\]
where Re stands for the real part. 
It is not hard to see that, for all $\mu_z\ge 0$ and $\lambda_{xy}\in\bC$,
\[
\inf_{t\in\bR,\; X\succeq 0} L(t, X, \mu, \lambda)
\]
is a lower bound on the optimal value of~\rf(eqn:dualGamma2Scalar).
Rearranging, we have
\[
L(t, X, \mu, \lambda) = 2\mathrm{Re}\sB[\sum_{x\in\cD_1, y\in\cD_2} \lambda_{xy}a_{xy}] + t\sB[1 - \sum_{z\in\ctD} \mu_z ] + \ip <X, W(\lambda,\mu)>,
\]
where $W(\lambda,\mu)$ is the block-matrix of the same form as $X$, defined by
\begin{equation}
\label{eqn:W}
\sA[W(\lambda,\mu)]_{xy} = \begin{cases}
-\lambda_{xy} \Delta_{xy},& \text{if $x\in\cD_1$ and $y\in\cD_2$;} \\
-(\lambda_{xy} \Delta_{xy})^*,& \text{if $x\in\cD_2$ and $y\in\cD_1$;} \\
\mu_x I, & \text{if $x=y$;}\\
0, & \text{otherwise.}
\end{cases}
\end{equation}
In particular, $W$ is Hermitian.  
Thus, we see that the following dual problem gives a lower bound on the optimal value of~\rf(eqn:dualGamma2Scalar):
\begin{subequations}
\label{eqn:d2Gamma2Scalar}
\begin{alignat}{2}
&\mbox{\rm maximise} &\quad& 2\mathrm{Re}\sB[ \sum\nolimits_{x\in\cD_1, y\in\cD_2} \lambda_{xy}a_{xy} ]\label{eqn:d2Gamma2ScalarObj}\\
& \mbox{\rm subject to}&& \sum\nolimits_{z\in\ctD} \mu_z = 1\;;\\
&&& W(\lambda,\mu)\succeq 0\;; \label{eqn:d2Gamma2ScalarW}\\
&&& \mu_z\ge 0,\quad \lambda_{xy}\in\bC \qquad \text{for all $z\in\ctD$, $x\in\cD_1$ and $y\in\cD_2$.}
\end{alignat}
\end{subequations}
By Slater's condition, as~\rf(eqn:dualGamma2Scalar) is convex and strictly feasible, the optimisation problems~\rf(eqn:dualGamma2Scalar) and~\rf(eqn:d2Gamma2Scalar) have equal optimal values.

We can simplify~\rf(eqn:d2Gamma2Scalar).  Let $u\in\bC^{\cD_1}$ and $v\in\bC^{\cD_2}$ be vectors satisfying $|u\elem[x]|^2 = \mu_x$ and $|v\elem[y]|^2 = \mu_y$ for all $x$ and $y$.
As $W(\lambda,\mu)\succeq 0$, we may assume that $\lambda_{xy}=0$ whenever $\mu_x\mu_y=0$.
Define $\gamma_{xy} = \lambda_{xy}/(u\elem[x]^* v\elem[y])$, where $0/0=0$.
Let us form the $\cD_1\times\cD_2$ matrices $A = (a_{xy})$ and $\Gamma = (\gamma_{xy})$.
Taking the Hadamard product of~\rf(eqn:d2Gamma2ScalarW) with an appropriate rank-1 positive-semidefinite matrix (and treating $z\in\ctD$ with $\mu_z=0$ separately), we have that~\rf(eqn:d2Gamma2ScalarW) is equivalent to
\[
\begin{pmatrix}
	I & \Gamma\circ\Delta \\
	(\Gamma\circ\Delta)^* & I 
\end{pmatrix}\succeq 0,
\]
which in turn is equivalent to $\|\Gamma\circ\Delta\|\le 1$.

On the other hand, the objective value~\rf(eqn:d2Gamma2ScalarObj) is equal to $2\mathrm{Re}\sA[u^*(\Gamma\circ A)v]$ subject to $\|u\|^2 + \|v\|^2 = 1$.
The maximum of this expression is $\|\Gamma\circ A\|$.
Thus, we get the dual formulation of \rf(prp:gamma2dual).

\subsection{Derivation of~\rf(eqn:relationApproximate)}
\label{app:relation}
Here we assume that $\eps>0$.  The case $\eps=0$ is considered in \rf(app:relationExact).
In our derivation, we use $\Delta_{xy} = I_\cX - O_x^*O_y$.  By \rf(rem:alternative), this choice is equivalent to $\Delta_{xy} = O_x - O_y$, and it also does not affect~\rf(eqn:relationApproximateCondition).  This choice has an important property $\Delta_{xy} = \Delta^*_{yx}$.

We transform the optimisation problem~\rf(eqn:relationDual) in a way similar to \rf(app:Gamma2Scalar).  Let $\cD'$ be a disjoint copy of $\cD$.  For $x\in\cD$, let $x'$ denote the corresponding element of $\cD'$, and let $\ctD = \cD\cup \cD'$.
Using a transformation similar to \rf(app:Gamma2Scalar), we see that~\rf(eqn:relationDual) is equivalent to
\begin{subequations}
\label{eqn:dualRelation}
\begin{alignat}{3}
&\mbox{\rm minimise} &\quad& t \\
& \mbox{\rm subject to}&&  \tr X_{zz} \le t &\quad& \text{for all $z\in\ctD$\;;} \\
&&& 1-\sum\nolimits_{a\in[m]}Y\elem[x,y]  = \ip<X_{xy'}, \Delta_{xy}> && \text{for all $x,y\in\cD$\;;}\\
&&& \sum\nolimits_{a\notin r(x)} Y_a\elem[x,x] \le \eps &&\text{for all $x\in\cD$\;;}\\
&&& t\in\bR,\quad X\succeq 0,\quad Y_a\succeq 0 &&\text{for all $a\in[m]$;}
\end{alignat}
\end{subequations}
where the optimisation is over $\cD\times\cD$ positive-semidefinite matrices $Y_a$, and $X$ is of the same form as in~\rf(app:Gamma2Scalar).  That is, $X$ is a $\ctD\times\ctD$ block matrix with each block acting on $\cX$.

The Lagrangian is equal to 
\begin{multline*}
t + \sum_{z\in\ctD} \mu_z (\tr X_{zz} - t) \\+ 
\sum_{x,y\in\cD} 2\mathrm{Re}\sC[{\lambda_{xy} \sB[ 1 - \sum_{a\in[m]} {Y_a\elem[x,y]} - \ip<X_{xy'},\Delta_{xy}>]}]
+ \sum_{x\in\cD} 2\eta_x \sC[ \sum_{a\notin r(x)} {Y_a\elem[x,x]} - \eps ],
\end{multline*}
where $\mu_z\ge 0$, $\eta_x \ge 0$, and $\lambda_{xy}\in\bC$.
After rearrangement, we get
\[
2\sum_{x,y\in\cD} \lambda_{xy} -2\eps\sum_{x\in\cD} \eta_x + t\sB[1 - \sum_{z\in\ctD} \mu_z ] + \ipA <X, W(\lambda,\mu)> + \sum_a \ipA<Y_a, 2 \mathrm H\circ \mathrm E_a -\Lambda - \Lambda^*>.
\]
Here, $W(\lambda,\mu)$ is defined as in~\rf(eqn:W), $\mathrm H = \mathrm{diag}(\eta_x)$, $\mathrm E_a = \mathrm{diag}(1_{a\notin r(x)})_{x\in\cD}$ and $\Lambda = (\lambda_{xy})_{x,y\in\cD}$.
This gives the following dual problem:
\begin{subequations}
\label{eqn:d2Relation}
\begin{alignat}{3}
&\mbox{\rm maximise} &\quad& 2\s[ \sum\nolimits_{x, y\in\cD} \mathrm{Re}\,\lambda_{xy} - \eps \sum\nolimits_{x\in\cD} \eta_x] \\
& \mbox{\rm subject to}&& \sum\nolimits_{z\in\ctD} \mu_z = 1;\qquad  W(\lambda,\mu)\succeq 0;\\
&&& \Lambda + \Lambda^* \preceq 2\; \mathrm H \circ \mathrm E_a &\quad& \text{for all $a\in[m]$;} \\
&&& \mu_z\ge 0,\quad \eta_x\ge 0,\quad \lambda_{xy}\in\bC && \text{for all $z\in\ctD$ and $x,y\in\cD$.}
\end{alignat}
\end{subequations}
By Slater's condition, the optimal values of~\rf(eqn:dualRelation) and~\rf(eqn:d2Relation) are equal.

Recall that $\Delta_{xy} = \Delta^*_{yx}$ for all $x,y\in\cD$.
This means that if $(\mu, \eta, \lambda)$ is a feasible solution to~\rf(eqn:d2Relation), then so is $(\mu', \eta, \lambda'$), where $\mu_x' = \mu_{x'}$, $\mu'_{x'} = \mu_x$ and $\lambda_{xy}' = \lambda_{yx}^*$ for all $x,y\in \cD$.  Moreover, the corresponding objective values are equal.  Since a convex combination of the two is still a feasible solution, we may assume, without loss of generality, that $\lambda_{xy} = \lambda_{yx}^*$ and $\mu_x = \mu_{x'}$ for all $x,y\in\cD$.  In particular, $\Lambda$ is Hermitian.

As in \rf(app:Gamma2Scalar), let $u\in\bC^\cD$ be a vector satisfying $|u\elem[x]|^2 = \mu_x$ for all $x\in\cD$.  
Again, we define $\gamma_{xy} = \lambda_{xy}/(u\elem[x]^*u\elem[y])$ and $\nu_x = \eta_x/\mu_x$.  A transformation similar to the one in \rf(app:Gamma2Scalar) shows that~\rf(eqn:d2Relation) is equivalent to~\rf(eqn:relationApproximate), where $\mathrm N = \mathrm{diag} (\nu_x)$.

\subsection{Derivation of~\rf(eqn:relationExact)}
\label{app:relationExact}
The derivation is similar to~\rf(app:relation).  This time, instead of~\rf(eqn:dualRelation), we have the following optimisation problem
\begin{subequations}
\label{eqn:dualExact}
\begin{alignat}{3}
&\mbox{\rm minimise} &\quad& t \\
& \mbox{\rm subject to}&&  \tr X_{zz} \le t &\quad& \text{for all $z\in\ctD$\;;} \\
&&& 1-\sum\nolimits_{a\in r(x)\cap r(y)}Y\elem[x,y]  = \ip<X_{xy'},\Delta_{xy}> && \text{for all $x,y\in\cD$\;;}\\
&&& t\in\bR,\quad X\succeq 0,\quad Y_a\succeq 0 &&\text{for all $a\in[m]$;}
\end{alignat}
\end{subequations}
with the difference that $Y_a$ is an $r^{-1}(a)\times r^{-1}(a)$ matrix.
The Lagrangian equals
\[
t + \sum_{z\in\ctD} \mu_z (\tr X_{zz} - t) + \sum_{x,y\in\cD} 2\mathrm{Re}\sC[{\lambda_{xy} \sB[ 1 - \sum_{a\in r(x)\cap r(y)} {Y_a\elem[x,y]} - \ip<X_{xy'},\Delta_{xy}>]}]\;,
\]
where $\mu_z\ge 0$, and $\lambda_{xy}\in\bC$.
After rearrangement, we get
\[
2\sum_{x,y\in\cD} \mathrm{Re}\,\lambda_{xy} + t\sB[1 - \sum_{z\in\ctD} \mu_z ] + \ip <X, W(\lambda,\mu)> + \sum_a \ipA<Y_a, -\Lambda_a  - \Lambda^*_a>,
\]
where $\Lambda_a = \Lambda\elem[r^{-1}(a), r^{-1}(a)]$.  Using the same transformation as in \rf(app:relation), we get that the optimal values of~\rf(eqn:dualExact) and~\rf(eqn:relationExact) are equal.

\subsection{Proof of \rf(thm:averageError)}
\label{app:average}
Using the same construction as in \rf(app:relation), we get the following optimisation problem:
\begin{subequations}
\label{eqn:dualAverage}
\begin{alignat}{3}
&\mbox{\rm minimise} &\quad& t \\
& \mbox{\rm subject to}&&  \tr X_{zz} \le t &\quad& \text{for all $z\in\ctD$\;;} \\
&&& 1-\sum\nolimits_{a\in[m]}Y\elem[x,y]  = \ip<X_{xy'},\Delta_{xy}> && \text{for all $x,y\in\cD$\;;}\\
&&& \sum_{x\in\cD} p_x \sum\nolimits_{a\notin r(x)} Y_a\elem[x,x] \le \eps\\
&&& t\in\bR,\quad X\succeq 0,\quad Y_a\succeq 0 &&\text{for all $a\in[m]$.}
\end{alignat}
\end{subequations}
This gives the dual
\begin{subequations}
\label{eqn:d2Average}
\begin{alignat}{3}
&\mbox{\rm maximise} &\quad& 2\s[ \sum\nolimits_{x, y\in\cD} \mathrm{Re}\,\lambda_{xy} - \eps \eta] \\
& \mbox{\rm subject to}&& \sum\nolimits_{z\in\ctD} \mu_z = 1;\qquad  W(\lambda,\mu)\succeq 0;\\
&&& \Lambda + \Lambda^* \preceq 2\; \mathrm \eta P \circ \mathrm E_a &\quad& \text{for all $a\in[m]$;} \\
&&& \mu_z\ge 0,\quad \eta\ge 0,\quad \lambda_{xy}\in\bC && \text{for all $z\in\ctD$ and $x,y\in\cD$.}
\end{alignat}
\end{subequations}
Again, we may assume that $\lambda_{xy}=\lambda_{yx}^*$ and $\mu_x = \mu_{x'}$.  Using transformations similar to \rf(app:relation), we get that~\rf(eqn:d2Average) is equivalent to~\rf(eqn:averageError).
\end{document}

%% file: _paper.tex
\input{\relativepath _MainMacros}

\newcommand{\CrossRef}[2]{#1~\ref{#2}}

\bigone{\newtheorem{thm}{Theorem}[chapter]}
\smallone{\newtheorem{thm}{Theorem}}

\newcommand{\maketheorem}[2]{
\newtheorem{#1}[thm]{#2}
\expandafter\def \csname ref#1\endcsname ##1{\CrossRef{#2}{#1:##1}}
}

\maketheorem{lem}{Lemma}
\maketheorem{prp}{Proposition}
\maketheorem{cor}{Corollary}
\maketheorem{clm}{Claim}
\maketheorem{fact}{Fact}

\theoremstyle{definition}
\maketheorem{rem}{Remark}
\maketheorem{obs}{Observation}
\maketheorem{defn}{Definition}
\maketheorem{exm}{Example}
\maketheorem{assump}{Assumption}

\def \rf(#1:#2){\csname ref#1\endcsname{#2}}
\def \rfitem(#1@#2){\rf(#2)(\ref{#1@#2})}
\def \rfitemE(#1@#2){\ref{#2}(\ref{#1@#2})}

\def\mycommand#1#2{
\expandafter\newcommand \csname#1\endcsname {#2}%
}
\def\remycommand#1#2{
\expandafter\renewcommand \csname#1\endcsname {#2}%
}


%% file: _MainMacros.tex
\usepackage{amsfonts,amssymb,amsmath,amsthm,latexsym}
\usepackage{array,xspace}

\newcommand{\ignore}[1]{}

\newcommand{\ii}{\mathsf{i}}
\newcommand{\ee}{\mathsf{e}}
\newcommand{\eps}{\varepsilon}

\newcommand{\tr}{\mathop{\mathrm{tr}}}

\newcommand{\etal}{{\em et al.}\xspace}

\def\makeletter#1{%
\expandafter \newcommand \csname b#1\endcsname {\mathbb{#1}}%
\expandafter \newcommand \csname c#1\endcsname {\mathcal{#1}}%
\expandafter \newcommand \csname t#1\endcsname {\widetilde{#1}}%
\expandafter \newcommand \csname ct#1\endcsname {\widetilde{\mathcal{#1}}}%
}
\def\makeletters(#1#2){\makeletter#1\ifx#2.\else\makeletters(#2)\fi}
\makeletters(QWERTYUIOPASDFGHJKLZXCVBNM.)

\def\makeSkob#1#2#3{%
\def\LLL{\left} \def\RRR{\right}
\expandafter \edef \csname #1\endcsname #2##1#3{\SkobInner}
\def\LLL{\bigl} \def\RRR{\bigr}
\expandafter \edef \csname #1A\endcsname #2##1#3{\SkobInner}
\def\LLL{\Bigl} \def\RRR{\Bigr}
\expandafter \edef \csname #1B\endcsname #2##1#3{\SkobInner}
\def\LLL{\biggl} \def\RRR{\biggr}
\expandafter \edef \csname #1C\endcsname #2##1#3{\SkobInner}
\def\LLL{\Biggl} \def\RRR{\Biggr}
\expandafter \edef \csname #1D\endcsname #2##1#3{\SkobInner}
\def\LLL{} \def\RRR{}
\expandafter \edef \csname #1O\endcsname #2##1#3{\SkobInner}
}

\def\SkobInner{\LLL(##1\RRR)} \makeSkob{s}[]
\def\SkobInner{\LLL[##1\RRR]} \makeSkob{sk}[]
\def\SkobInner{\LLL\lbrace##1\RRR\rbrace} \makeSkob{sfig}{}{}
\def\SkobInner{\LLL\lfloor##1\RRR\rfloor} \makeSkob{floor}[]
\def\SkobInner{\LLL\lceil##1\RRR\rceil} \makeSkob{ceil}[]
\def\SkobInner{\LLL\langle##1\RRR\rangle} \makeSkob{ip}<>
\def\SkobInner{\LLL|##1\RRR\rangle} \makeSkob{ket}|>
\def\SkobInner{\LLL|##1\RRR|} \makeSkob{abs}||
\def\SkobInner{\LLL\|##1\RRR\|} \makeSkob{norm}||
\def\SkobInner{\LLL\|##1\RRR\|_{\noexpand\mathrm F}} \makeSkob{normFrob}||
\def\SkobInner{\LLL\|##1\RRR\|_{\noexpand\mathrm{tr}}} \makeSkob{normtr}||

\newcommand{\midA}{\mathbin{\bigl|}}
\newcommand{\midB}{\mathbin{\Bigl|}}

\def \elem[#1]{[\![#1]\!]}
\def \bigfrac#1/{\left.#1\right/}
\def \bigfracR/#1.{\left/#1\right.}

\newcommand{\pfstart}{\begin{proof}} 
\newcommand{\pfsketch}{\begin{proof}[Proof sketch]}
\newcommand{\pfend}{\end{proof}} 
\newcommand{\itemstart}{\begin{itemize}\itemsep0pt}
\newcommand{\itemend}{\end{itemize}}
\newcommand{\descrstart}{\begin{description}\itemsep0pt}
\newcommand{\descrend}{\end{description}}
\newcommand{\enumstart}{\begin{enumerate}\itemsep0pt}
\newcommand{\enumend}{\end{enumerate}}

\newcommand{\negmedskip}{\vspace{-\medskipamount}}